\documentclass[aps,pra,twocolumn,groupedaddress,longbibliography,nobibnotes]{revtex4-1}
\usepackage{amsmath}
\usepackage{amssymb}
\usepackage{mathrsfs}
\usepackage[pdftex,colorlinks=true,urlcolor=blue,linkcolor=blue,citecolor=blue,breaklinks=true]{hyperref}
\usepackage{graphicx}
\usepackage{float}
\usepackage{dsfont}
\usepackage{multibib}
\newcommand{\push}{\hspace{0.05cm}}
\newcommand{\pull}{\hspace{-0.05cm}}

\makeatletter
\def\l@subsection{%
 \l@@sections{section}{subsection}% Implicit #3#4
}
\def\l@f@subsection{%
 \addpenalty{\@secpenalty}%
 \addvspace{0.5em plus\p@}%
}
\def\l@subsubsection#1#2{}
\makeatother

\begin{document}

\title{Out-of-equilibrium steady states of a locally driven lossy qubit array}

\author{Shovan Dutta}
\email[E-mail: ]{sd843@cam.ac.uk}
\author{Nigel R. Cooper}
\email[E-mail: ]{nrc25@cam.ac.uk}
\affiliation{T.C.M. Group, Cavendish Laboratory, University of Cambridge, JJ Thomson Avenue, Cambridge CB3 0HE, United Kingdom\looseness=-1}

\date{\today}

\begin{abstract}
We find a rich variety of counterintuitive features in the steady states of a qubit array coupled to a dissipative source and sink at two arbitrary sites, using a master equation approach. We show there are setups where increasing the pump and loss rates establishes long-range coherence. At sufficiently strong dissipation, the source or sink effectively generates correlation between its neighboring sites, leading to a striking density-wave order for a class of ``resonant'' geometries. This effect can be used more widely to engineer nonequilibrium phases. We show the steady states are generically distinct for hard-core bosons and free fermions, and differ significantly from the ones found before in special cases. They are explained by generally applicable ansatzes for the long-time dynamics at weak and strong dissipation. Our findings are relevant for existing photonic setups.
%We find a rich variety of counterintuitive features in the steady states of a qubit array coupled to a dissipative source and sink at two arbitrary sites, using a master equation approach. We show there are setups where increasing the pump and loss rates establishes long-range coherence. At strong dissipation, the source/sink effectively generates correlation between its neighboring sites, leading to characteristic density waves for a certain class of arrangements satisfying a resonance condition. Even at weak dissipation the steady state is generically nonthermal, and distinct for free fermions and hard-core bosons. These results are explained by a general ansatz for the long-time dynamics, which would be useful in other systems. Our findings are relevant for existing photonic setups.
\end{abstract}

\maketitle

\section{\label{intro}Introduction}
Environmental decoherence has long been seen as an unavoidable roadblock to stabilizing quantum phases for long periods of time \cite{Schlosshauer2019}. However, rapid advances in cooling and trapping techniques over the last decades have led to experimental platforms where the coupling to the environment can be controlled and even engineered to an unprecedented degree \cite{Mueller2012}. As several studies have shown, such tailored dissipation can be used to prepare novel quantum states \cite{Diehl2011, Lin2013, Carr2013}. The competition between Hamiltonian dynamics and incoherent dissipation can produce feature-rich steady states with no analogue in equilibrium condensed matter \cite{Sieberer2016}. Understanding these nonequilibrium phases is of fundamental interest \cite{Kordas2015}, with potential applications in quantum computing \cite{Verstraete2009}.

A prototypical experimental setup for exploring such states is a one-dimensional (1D) array of qubits coupled to local reservoirs. In particular, the qubits can be realized by hard-core bosons on a lattice \cite{Cazalilla2011}, or equivalently, a spin-1/2 XY chain, and the reservoir(s) can be designed to inject or remove a particle (or flip spin) at a given site, as in Ref.~\cite{Ma2019}. Theoretical studies modeling the resulting dynamics have focused almost exclusively on the cases where the pump and loss occur at the ends of the chain. Then the system can be reduced to free fermions \cite{Prosen2008}, enabling special analytical approaches that have been used to examine nonequilibrium transport \cite{Prosen2008, Prosen2009, Znidaric2010, Znidaric2011, Prosen2011a, Znidaric2011a, Kos2017} and phase transitions \cite{Prosen2008quantum, Znidaric2011solvable, Banchi2014}. However, without additional Zeeman fields, the steady state for end drives is rather featureless, with no long-range order \cite{Prosen2008quantum, vznidarivc2010matrix}. On the other hand, a recent work showed that for pump and loss at the center, there are multiple steady states with long-range coherence, that are distinct from free fermions and arise from a dynamical symmetry \cite{Dutta2020}. Such disparate results beg the question of what happens for generic pump-loss configurations, where pump and loss are neither both at the end nor both at the center. 

Here we characterize the steady states for generic setups with a single pump and a single loss site, finding several counterintuitive features which can be probed in already existing platforms \cite{Ma2019}. In Sec.~\ref{corrvsrate}, we use perturbation theory supported by numerics to show there are dipole-like arrangements where long-range coherence is induced by increasing dissipation. In Sec.~\ref{nonthermal}, we show the steady state is generically nonthermal even at weak dissipation, contrary to what is known for symmetric setups \cite{Buca2014}. Further, hard-core bosons and free fermions can form qualitatively distinct steady-state correlations, although their density profiles are always reflection symmetric. These attributes are explained by a simple product ansatz of the single-particle modes. In Sec.~\ref{resonance}, we find that at strong dissipation, the chain is generally divided into a filled and an empty segment separated by a high-entropy bulk. These segments are coupled by the source or sink which effectively produces correlation. Whenever two modes in neighboring segments come into resonance, this effect leads to striking density waves and long-range order. This is a geometric effect and can be generalized to multiple sources and sinks. We explain the oscillations by a modified ansatz, finding they are more robust in free fermions than in hard-core bosons.

%in the strongly dissipative Zeno regime, the pump and loss generally divide the chain into a filled and an empty segment separated by a high-entropy region in the middle. However, this is dramatically altered whenever two modes in neighboring segments come into resonance. Then particles can virtually hop between segments, leading to characteristic density oscillations and long-range order. This is a purely geometric effect and can be generalized to multiple sources and sinks. We explain the oscillations by a modified ansatz that incorporates correlation between the resonant modes, finding they are more robust in free fermions than in hard-core bosons.

These results highlight surprising phenomena that can arise in open many-body settings, elucidating differences between hard-core bosons and free fermions in 1D \cite{Malo2018}. Our ansatzes apply to more general forms of dissipation, and reduce the numerical cost to linear in system size.

\section{\label{model}Model and known special cases}
We consider strongly interacting bosons on a 1D lattice in the hard-core limit \cite{Cazalilla2011}, described by the Hamiltonian
\begin{equation}
\hat{H} = -\hbar J \push\sum\nolimits_{i=1}^{L-1} \big(\hat{b}_{i}^{\dagger} \hat{b}_{i+1} + \hat{b}_{i+1}^{\dagger} \hat{b}_{i}\big)\;,
\label{bosonhamil}
\end{equation}
where $\smash{\hat{b}_i^{\dagger}}$ is the boson creation operator, $J$ is the hopping amplitude, and $L$ is the number of sites. The hard-core constraint is encoded in the relation $\smash{\hat{b}_i^{\dagger 2} \pull= 0}$, which means no two bosons can occupy the same site. This leads to the commutation rules $\smash{[\hat{b}_i,\hat{b}_j]=0}$ and $\smash{[\hat{b}_i,\hat{b}_j^{\dagger}] = (-1)^{\hat{n}_i} \delta_{ij}}$, where $\smash{\hat{n}_i:=\hat{b}_i^{\dagger}\hat{b}_i}$ is the local occupation. Such a system is equivalent to a spin-1/2 XX chain, and has been realized with cold atoms in optical lattices \cite{Paredes2004, Stoeferle2004, Preiss2015} and microwave photons in nonlinear resonators \cite{Ma2019}. The Hamiltonian is reduced to free fermions by the Jordan-Wigner map
\begin{equation}
\hat{f}_j = (-1)^{\sum_{i<j} \hat{n}_i} \hat{b}_j\;,
\label{JordanWigner}
\end{equation}
where $\smash{\hat{f}_j}$ are the free fermion operators. The transformed Hamiltonian simply reads $\hat{H} = -\hbar J\sum_i \big(\hat{f}_i^{\dagger} \hat{f}_{i+1} + \text{H.c.}\big)$.

%We assume the system is coupled to local Markovian reservoirs that inject bosons at site $p$, if it is empty, and remove bosons from site $q$, if it is occupied. Such sources and sinks have been engineered via sideband transitions in microwave circuits \cite{Ma2019} and electron beams in optical traps \cite{Barontini2013}. Upon tracing out the environment, the density operator $\hat{\rho}$ is governed by the master equation \cite{Lindblad1976, Gorini1976, breuer2002theory}

The system of bosons is coupled to bosonic reservoirs that inject particles at a site $p$, if it is empty, and remove bosons from a site $q$, if it is occupied. Such sources and sinks have been engineered using transmon qubits in microwave circuits \cite{Ma2019} and electron beams in optical traps \cite{Barontini2013}. We assume the reservoirs are Markovian, i.e., they relax to equilibrium much faster than the system dynamics and the coupling, which is standard for these setups \cite{Daley2014}. Upon tracing out the environment, the density operator $\hat{\rho}$ is governed by the master equation \cite{Kordas2015, Daley2014, Lindblad1976, Gorini1976, breuer2002theory}
\begin{equation}
\frac{d\hat{\rho}}{dt} = \mathcal{L}\hat{\rho} := -\frac{{\rm i}}{\hbar}\push [\hat{H},\hat{\rho}] \push + \sum\nolimits_{\alpha}\pull \hat{L}_{\alpha} \hat{\rho}\hat{L}_{\alpha}^{\dagger} - \frac{1}{2} \{\hat{L}_{\alpha}^{\dagger} \hat{L}_{\alpha}, \hat{\rho}\}\;,
\label{mastereqn}
\end{equation}
where we have two Lindblad operators modeling the dissipation, $\smash{\hat{L}_+ := \sqrt{\gamma_+}\push\hat{b}_p^{\dagger}}$ and $\smash{\hat{L}_- := \sqrt{\gamma_-}\push\hat{b}_q}$, $\gamma_{\pm}$ being the pump and loss rates, respectively. Note that Eq.~\eqref{JordanWigner} does not, in general, reduce the full dynamics to free fermions with (local) pump and loss. Instead, the dissipation mediates nonlocal interactions between the fermions.

The above system has been studied most widely when the source and sink are at opposite ends, i.e., $p=1$ and $q=L$. Then the only term in Eq.~\eqref{mastereqn} that differs from the free-fermion case is $\smash{\hat{L}_- \hat{\rho} \hat{L}_-^{\dagger} = \gamma_- \hat{f}_L \hat{P} \hat{\rho} \hat{P} \hat{f}_L^{\dagger}}$, where $\smash{\hat{P}}$ is the total particle-number parity. Since $\hat{P}$ is conserved by the Hamiltonian, one can show the dynamics decouple into sectors with $\smash{\hat{P}\hat{\rho} \hat{P} = \pm \hat{\rho}}$. Thus, the Liouvillian $\mathcal{L}$ is quadratic in the free fermions. In such cases, the full solution can be found from the spectral properties of a $4L\times 4L$ matrix using quantization in the space of operators \cite{Prosen2008}. The steady state is identical to that of end-driven free fermions, characterized by a uniform bulk with short-range correlations \cite{vznidarivc2010matrix}, as shown in Fig.~\ref{knownplot}(a). We present a closed-form solution in the \hyperref[senddrivesol]{Supplement} \cite{supplement}.

\begin{figure}
\centering
\includegraphics[width=\columnwidth]{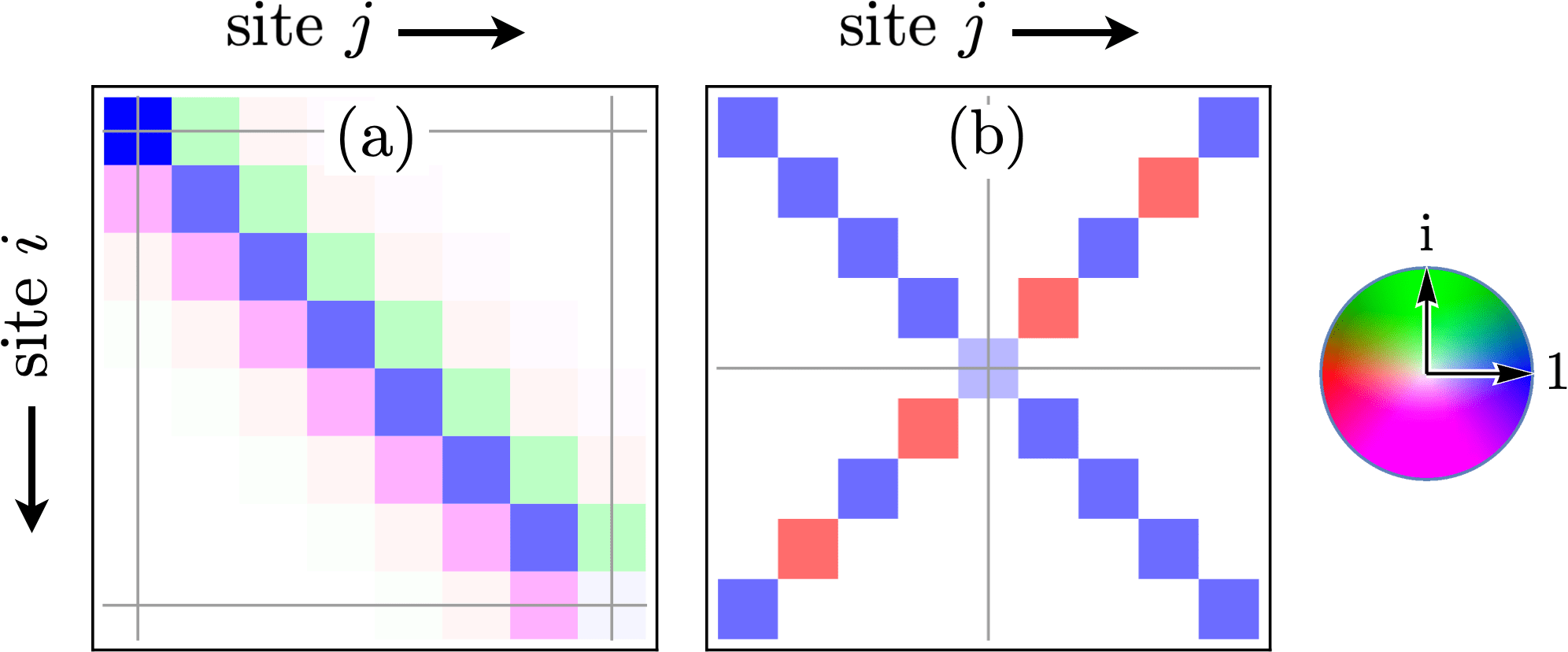}
\caption{\label{knownplot}Contrasting steady-state correlations $\smash{\langle\hat{b}_i^{\dagger}\hat{b}_j\rangle}$ of hard-core bosons on a 1D lattice subject to incoherent pump with rate $\gamma_+$ and loss with rate $\gamma_-$ (a) at opposite ends, with $\gamma_+ = \gamma_- = 10 J$, and (b) at the center, with $\gamma_-=4\gamma_+$.
}
\end{figure}

The end-driven case is in sharp contrast to the scenario where pump and loss both occur at the center site (for odd $L$), which we explored in a recent work \cite{Dutta2020}. Here, the system does not map onto free fermions. Instead, one has multiple steady states due to a symmetry operator $\hat{C} = -1/2+\sum_k \hat{f}_{L+1-k}^{\dagger} \hat{f}_{k}$, which splits the dynamics into $(L-1)/2$ sectors with varying degrees of entanglement. The symmetry stabilizes particle-hole pairs at reflection-symmetric sites $k$ and $\smash{\tilde{k}:=L+1-k}$, leading to steady states with long-range coherence, as shown in Fig.~\ref{knownplot}(b) for the maximally entangled sector. In comparison, free fermions have an exponentially large set of steady states, as all odd single-particle states vanish at the center.

Outside the above two scenarios, it is known that driving with both pump and loss at a generic site (not center) yields the product state $\smash{\hat{\rho}_0 \propto \otimes_i (\gamma_+ |1_i\rangle \langle 1_i| \hspace{-0.01cm}+\hspace{-0.01cm} \gamma_- |0_i\rangle \langle 0_i|)}$ \cite{Pizorn2013}, which is an infinite-temperature state with chemical potential $\mu=\ln(\gamma_+/\gamma_-)$. We find free fermions also have the same steady state, except when a single-particle state vanishes at the drive site, producing degeneracies.

\section{\label{corrvsrate}Dissipation induced long-range coherence}
\begin{figure}
\centering
\includegraphics[width=\columnwidth]{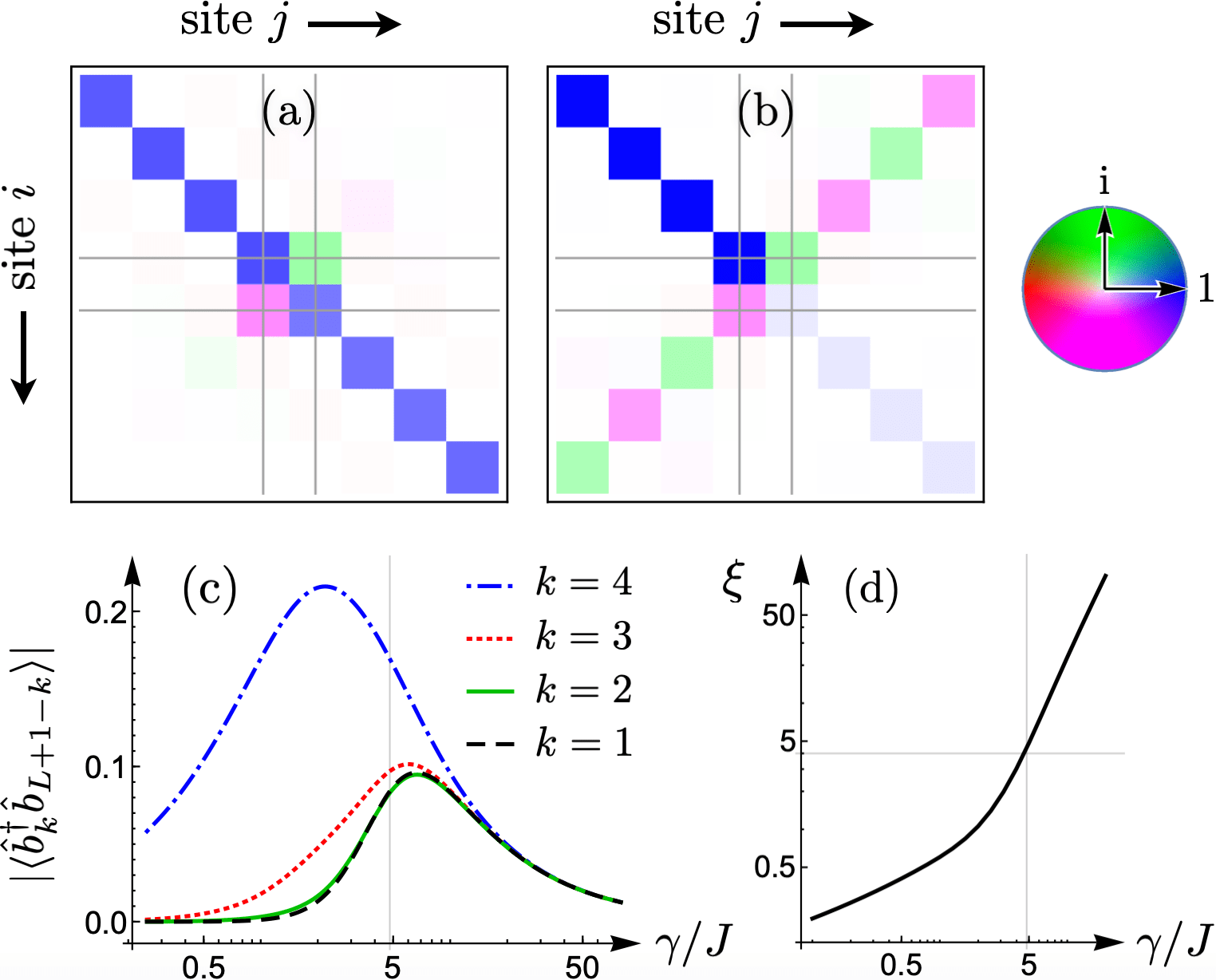}
\caption{\label{corrvsdissplot}Steady-state correlations $\smash{\langle\hat{b}_i^{\dagger}\hat{b}_j\rangle}$ for pump and loss at neighboring sites $p=L/2$, $q=L/2+1$ for $\gamma_+ = \gamma_- :=\gamma$ and $L=8$ at (a) weak dissipation, $\gamma =0.6J$, and (b) strong dissipation, $\gamma = 8J$. (c) Coherence between reflection-symmetric sites as a function of the pump/loss rate. (d) Correlation length $\xi$ obtained by fitting the coherences to an exponential.
}
\end{figure}

As described above, long-range order is absent for end drives, and restored for center drives by a special symmetry, irrespective of the pump/loss rate \cite{Dutta2020}. Here we find examples where long-range coherence is established by increasing dissipation. In particular, consider a ``dipole'' setup where pump and loss occur at two neighboring sites in the middle, i.e., $p=L/2$ and $q=L/2+1$, for even $L$. This can be seen as a center-drive analogue, but there is no strong symmetry \cite{Buca2012} and the steady state is unique. Using first-order perturbation theory at weak dissipation ($\gamma_{\pm} \ll J$), we find the steady state (see \hyperref[sdipolesol]{Supplement} \cite{supplement})
\begin{equation}
\hat{\rho}_{\text{w}} \approx \Big[ 1+ {\rm i}\push  \frac{\gamma_+ \pull + \gamma_-}{2J} \big(\hat{Q} - \hat{Q}^{\dagger}\big)\Big] \hat{\rho}_0\;,
\label{dipoleweak}
\end{equation}
where $\smash{\hat{Q} := \sum_{k=1}^{L/2} \hat{f}_{\tilde{k}}^{\dagger} \hat{f}_k}$, and $\hat{\rho}_0$ is the product state with uniform occupation $n_0 = \gamma_+/(\gamma_+ \pull + \gamma_-)$. The perturbation $\smash{\hat{Q}}$ is reminiscent of the symmetry operator $\smash{\hat{C}}$, and generates the antidiagonal correlations% in Ref.~\cite{Dutta2020}, and yields the off-diagonal correlations
\begin{equation}
\langle \hat{b}_k^{\dagger} \hat{b}_{\tilde{k}}\rangle_{\text{w}} \approx {\rm i}\push \frac{\gamma_+ \gamma_-}{2 J (\gamma_+ \pull + \gamma_-)} \bigg(\pull\frac{\gamma_+ \pull- \gamma_-}{\gamma_+ \pull+ \gamma_-}\pull\bigg)^{\pull L-2k}.
\label{dipoleweakcorr}
\end{equation}
Thus, at weak dissipation, the coherences decay exponentially with distance, and are limited to nearest neighbors ($k=L/2$) for $\gamma_+\pull=\gamma_-$, as shown in Fig.~\ref{corrvsdissplot}(a). Conversely, for strong dissipation, the steady state approaches $\hat{\rho}_{\text{step}}$ where all sites $i\leq L/2$ are filled and $i>L/2$ are empty. To first order in $J/\gamma_{\pm}$, we find (see \hyperref[sdipolesol]{Supplement} \cite{supplement}),
\begin{align}
&\hat{\rho}_{\text{s}} \approx \hat{\rho}_{\text{step}} + {\rm i} \push \frac{2 J}{\gamma_+ \pull + \gamma_-} \big(\hat{Q}\push\hat{\rho}_{\text{step}} - \text{H.c.}\big)\;,
\label{dipolestrong}\\
\text{and}\quad &\langle \hat{b}_k^{\dagger} \hat{b}_{\tilde{k}}\rangle_{\text{s}} \approx {\rm i}\push \frac{2 J}{\gamma_+ \pull + \gamma_-}\push (-1)^{L/2-k}\;.
\label{dipolestrongcorr}
\end{align}
As shown in Fig.~\ref{corrvsdissplot}(b), now the coherences span the entire system, similar to the center-driven case [Fig.~\ref{knownplot}(b)]. These results imply that long-range coherence is induced by increasing the pump/loss relative to tunneling, which is confirmed by exact numerics [Figs.~\ref{corrvsdissplot}(c) and \ref{corrvsdissplot}(d)]. The correlation length grows monotonically with dissipation, exceeding the system size for $\gamma_{\pm}\gtrsim 5J$. In contrast, for end drives [Fig.~\ref{knownplot}(a)], coherences are limited to nearest neighbors at both weak and strong dissipation, yielding (see \hyperref[senddrivesol]{Supplement} \cite{supplement})
\begin{equation}
\langle\hat{b}_j^{\dagger} \hat{b}_{j+1}\rangle_{\text{w}} \approx {\rm i}\push \frac{\gamma_+ \gamma_-}{2 J (\gamma_+ \pull + \gamma_-)}\push ;\; \langle\hat{b}_j^{\dagger} \hat{b}_{j+1}\rangle_{\text{s}} \approx {\rm i}\push \frac{2 J}{\gamma_+ \pull + \gamma_-}\;.
\end{equation}

Numerically, we find similar results in ``dipole'' setups with $q=p+1$, whenever $p$ divides $L-p$ or vice versa (see \hyperref[sotherdipoles]{Supplement} \cite{supplement} for examples). In Sec.~\ref{resonance}, we discuss more general scenarios where such geometric resonances stabilize long-range order in the Zeno limit.%long-range order survives in the Zeno limit but not at weak dissipation.

\section{\label{nonthermal}Nonthermal steady states at weak dissipation}
\begin{figure}[b]
\centering
\includegraphics[width=\columnwidth]{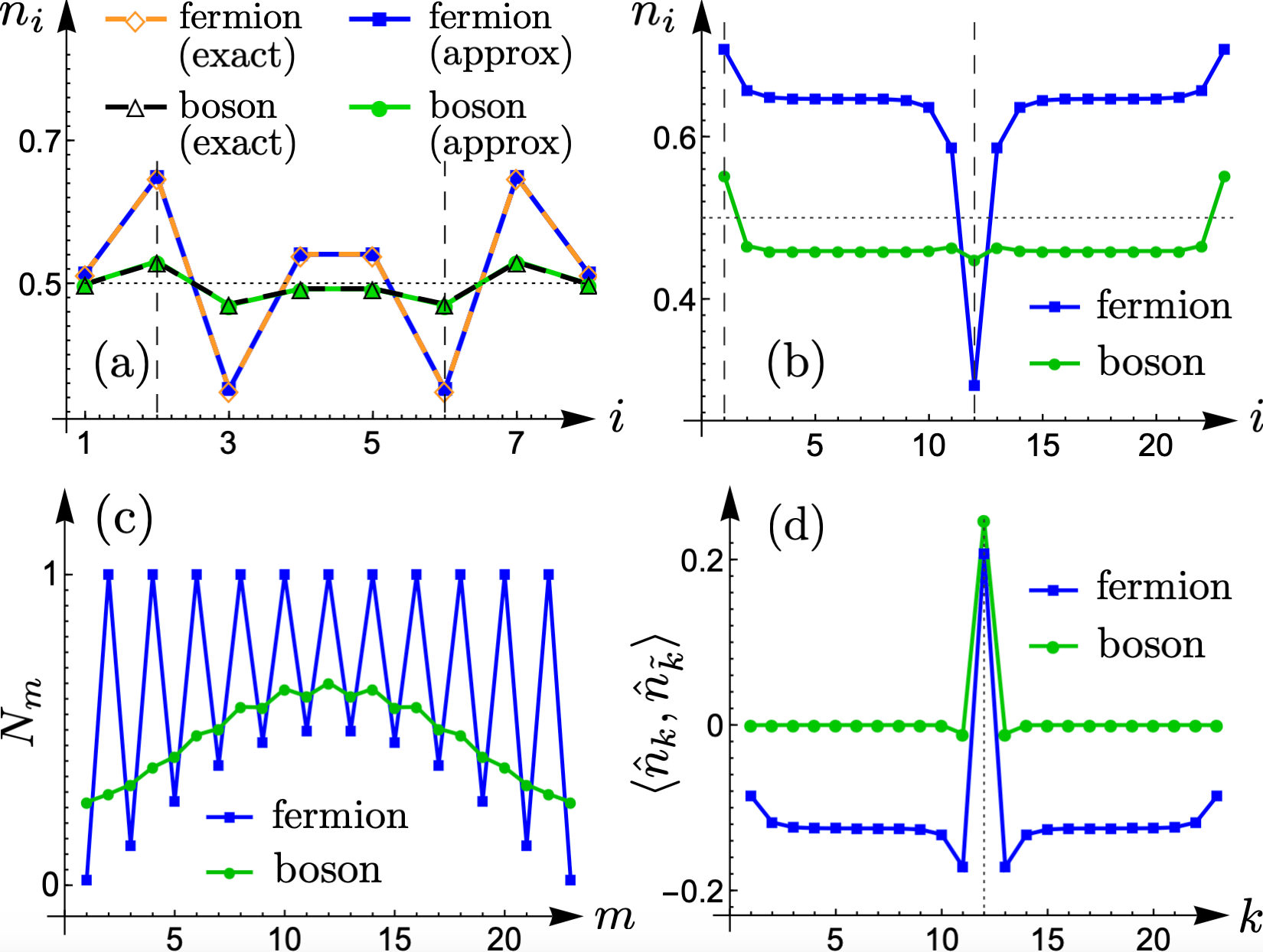}
\caption{\label{weakdissplot}Steady-state density $n_i$ of free fermions and hard-core bosons in the weak-coupling limit with $\gamma_+ \hspace{-0.06cm}=\hspace{-0.02cm} \gamma_-$ for (a) $L=8$, $p=2$, $q=6$, from exact diagonalization and using a product-of-modes ansatz [Eqs.~\eqref{weakratefermion}--\eqref{weakrateboson}], and (b) $L=23$, $p=1$, $q=12$, using the ansatz. (c) Occupation of single-particle modes and (d) density correlation between sites $k$ and $\smash{\tilde{k}:=L+1-k}$ for the setup in (b), where $\smash{\langle \hat{n}_i, \hat{n}_j\rangle :=\langle \hat{n}_i \hat{n}_j\rangle - n_i n_j}$.
}
\end{figure}

For end drives as well as for the ``dipole'' setup considered in Sec.~\ref{corrvsrate}, the steady state approaches the uniform infinite-temperature state $\hat{\rho}_0$ in the weak-coupling limit. Here, the system has time to equilibrate between successive pump and loss events. Thus, one might expect the same steady state regardless of where those events occur. This is indeed true whenever the pump and loss act on reflection-symmetric sites \cite{Buca2014} or at the same site (except center) \cite{Pizorn2013}. However, we find those are the only setups where the conjecture holds. As shown in Figs.~\ref{weakdissplot}(a)--(b), the steady state is generically nonthermal, and different for free fermions and hard-core bosons, although they both have symmetric densities. These features can be understood by focusing instead on the single-particle modes which are unaffected by the Hamiltonian. They are given by unitary maps $\smash{\hat{F}_m = \sum_{j} c_{m,j} \hat{f}_j}$ where
\begin{equation}
c_{m,j} = \sqrt{2/(L\pull+\pull 1)} \push\sin\push[\pi m j /(L\pull+\pull 1)]\;,
\label{modes}
\end{equation}
and have energy $\varepsilon_m = -2\hbar J \cos\pull\big(\frac{\pi m}{L+1}\big)$, $m=1,\dots,L$. If the dissipation is small compared to the energy splitting, the modes become uncorrelated. Thus, we find the steady state is well approximated by a ``product-of-modes" form $\hat{\rho} \approx \prod_m (1\pull-\pull N_m) |0_m\rangle\langle 0_m| + N_m |1_m\rangle\langle 1_m|$, set by the mode occupations $N_m$. The lack of correlation explains why the density is symmetric. Using $\smash{\langle \hat{F}_m^{\dagger} \hat{F}_n \rangle \pull\approx \delta_{mn}N_m}$ yields $\smash{n_j:=\langle\hat{n}_j\rangle = \sum_m\pull N_m |c_{m,j}|^2 }$, which gives $n_{\tilde{j}} = n_{j}$ for any site $j$. (Recall that $\smash{\tilde{j} := L+1-j}$.) %, since $|c_{m,\tilde{j}}| \pull=\pull |c_{m,j}|$.

One can use the product-of-modes ansatz in Eq.~\eqref{mastereqn} to construct approximate rate equations for $N_m$. For free fermions, one finds (see \hyperref[srateeqsweak]{Supplement} \cite{supplement})
\begin{equation}
\dot{N}_m \approx \gamma_+ |c_{m,p}|^2 (1-N_m) - \gamma_- |c_{m,q}|^2 N_m\;,
\label{weakratefermion}
\end{equation}
i.e., the modes are uncoupled, with incoming and outgoing currents set by the weight of the respective mode at the pump and loss sites. To reach a uniform steady state such as $\smash{\hat{\rho}_0}$, one must have $|c_{m,p}| = |c_{m,q}|$ $\forall m$, which is satisfied iff $q=p$ or $q=\tilde{p}$. While this conclusion also holds for hard-core bosons, the rate equations are more complex as the string operator in Eq.~\eqref{JordanWigner} generates nonlinear coupling between the modes. Using $\smash{\langle \hat{F}_m^{\dagger} \hat{F}_{m^{\prime}}^{\dagger} \hat{F}_n \hat{F}_{n^{\prime}} \rangle \approx}$ $N_n N_{n^{\prime}} (\delta_{m,n^{\prime}} \delta_{m^{\prime},n} \pull- \delta_{m,n} \delta_{m^{\prime},n^{\prime}})$, we find (see derivation in the \hyperref[srateeqsweak]{Supplement} \cite{supplement})
\begin{align}
\nonumber\hspace{-0.235cm} \dot{N}_m\hspace{-0.03cm} \approx &\;\gamma_+\big[\hspace{0.02cm}\bar{N}_m\hspace{-0.02cm} \big(N_m |c_{m,p}|^2 \hspace{-0.04cm}+\hspace{-0.01cm} \bar{n}_p\big)\pull - \bar{n}_p \bar{\beta}^{(p)}_m\pull + |\kappa^{(p)}_m\pull -\hspace{-0.01cm}c_{m,p}|^2 \hspace{0.02cm}\big]\\
-\pull&\;\gamma_- \big[\hspace{0.02cm} N_m \big( \bar{N}_m |c_{m,q}|^2 \hspace{-0.04cm}+\hspace{-0.01cm} n_q\big)\pull - n_q \beta^{(q)}_m\pull + |\kappa^{(q)}_m|^2 \hspace{0.02cm}\big]\push,
\label{weakrateboson}
\end{align}
with notation $\bar{x} :=1- x$, $\smash{\beta^{(i)}_m :=\sum_n\pull N_n |\alpha^{(i)}_{m,n}|^2}$, $\smash{\kappa^{(i)}_m :=}$ $\sum_n\pull N_n \alpha^{(i)}_{n,m} c_{n,i}$, where $\alpha^{(i)}_{m,n} := \big(\sum_{j\geq i} \pull- \sum_{j<i}\pull\big)\push c_{m,j}^* c_{n,j}$. In numerical trials, Eq.~\eqref{weakrateboson} gives a unique steady state with $0\leq N_m \leq 1$ for all $m$.

Figure~\ref{weakdissplot}(a) shows the ansatz accurately describes the steady states in the weak-coupling limit, becoming exact for free fermions. The densities are peaked at the pump site and minimized at the loss site, with $n_q = 1-n_p$ for $\gamma_+ = \gamma_-$. The same is mirrored at sites $\tilde{p}$ and $\tilde{q}$, which explains why choosing $q=\tilde{p}$ \cite{Buca2014} or $q=p$ \cite{Pizorn2013} gives half filling at all sites. In general, the density fluctuations are significantly smaller for hard-core bosons due to strong interactions. There are also qualitative differences which persist to large system sizes, as shown in Figs.~\ref{weakdissplot}(b)--(d). Here one has pump at one end and loss at the center. For free fermions, all the odd modes (even $m$) are immune to loss and thus fully filled, which is not the case for hard-core bosons [Fig.~\ref{weakdissplot}(c)]. Thus, we find strikingly different densities and correlations. In particular, Fig.~\ref{weakdissplot}(d) shows that unlike bosons, fermions exhibit long-range order in the density correlations $\langle \hat{n}_i \hat{n}_j\rangle - n_i n_j \approx \delta_{ij}n_i - \smash{|\langle\hat{f}_i^{\dagger}\hat{f}_j\rangle|^2}$ \cite{supplement}, which holds more generally at weak dissipation. Also, note the free fermions have degenerate steady states if any of the modes vanishes at both pump and loss sites, whereas for hard-core bosons the steady state is unique except for center drive \cite{Dutta2020}. Equations~\eqref{weakratefermion} and \eqref{weakrateboson} apply for any quadratic Hamiltonian with nondegenerate spectrum, reducing the dynamics to $L$ rate equations.

\section{\label{resonance}Geometric resonance and long-range order in Zeno limit}
In Sec.~\ref{corrvsrate} we found, for pump and loss at neighboring sites, $q=p+1$, the steady state at strong dissipation approaches a step where sites 1 through $p$ are filled and sites $q$ through $L$ are empty [Eq.~\eqref{dipolestrong}]. This can be understood as follows. For $t\gtrsim 1/\gamma_{\pm}$, sites $p$ and $q$ are pinned at occupation 1 and 0, respectively. By tracing over this subspace, one can show the remaining sites are governed by a master equation with an effective Hamiltonian $\smash{\hat{H}_{\text{eff}}}$ and weak effective dissipation, as detailed in Ref.~\cite{Popkov2018} for more general systems. In our setup, $\smash{\hat{H}_{\text{eff}}}$ simply describes hopping in the disjoint segments 1 to $p-1$, $p+1$ to $q-1$, and $q+1$ to $L$. For $q=p+1$, the middle region is absent and the dissipation is given by Lindblad operators $\smash{\hat{L}^{\text{eff}}_+ \pull= \pull\sqrt{\Gamma_+} \hat{b}_{p-1}^{\dagger}}$ and $\smash{\hat{L}^{\text{eff}}_- \pull= \pull\sqrt{\Gamma_-}\hat{b}_{q+1}}$ where $\Gamma_{\pm} \pull := 4J^2\pull /\gamma_{\pm}$ (see \hyperref[seffstrong]{Supplement} \cite{supplement}). The former injects particles into the first segment until it is filled, and the latter removes all particles from the last segment. More generally,
\begin{equation}
\hat{L}^{\text{eff}}_+ \pull=\pull \sqrt{\Gamma_+}\big(\hat{b}_{p-1}^{\dagger}\pull+\hat{b}_{p+1}^{\dagger}\big),\; \hat{L}^{\text{eff}}_- \pull=\pull \sqrt{\Gamma_-} \big(\hat{b}_{q-1}\pull+\hat{b}_{q+1}\big)\push,
\label{effectiveLs}
\end{equation}
i.e., the source and sink induce correlated pump and loss at neighboring sites with rate $\Gamma_{\pm}$, coupling the segments. However, this is a second-order effect ($\Gamma_{\pm}/J \sim J^2 /\gamma_{\pm}^2$), and the steady state remains uncorrelated whenever the energy splitting between modes in adjacent segments is large compared to $\Gamma_{\pm}$, as we explain below. Then the mid region behaves like an end-driven qubit array, reaching a uniform product state with density $n^{(2)} \hspace{-0.03cm}= \Gamma_+/(\Gamma_+\pull + \Gamma_-)$. This is similar to $\hat{\rho}_0$ in Eq.~\eqref{dipoleweak} except $n^{(2)}$ decreases with $\gamma_+/\gamma_-$, a consequence of the Zeno effect \cite{misra1977zeno}. Thus, the steady state generically consists of fully filled and empty sites separated by a region of high entropy.

\begin{figure}[b]
\centering
\includegraphics[width=\columnwidth]{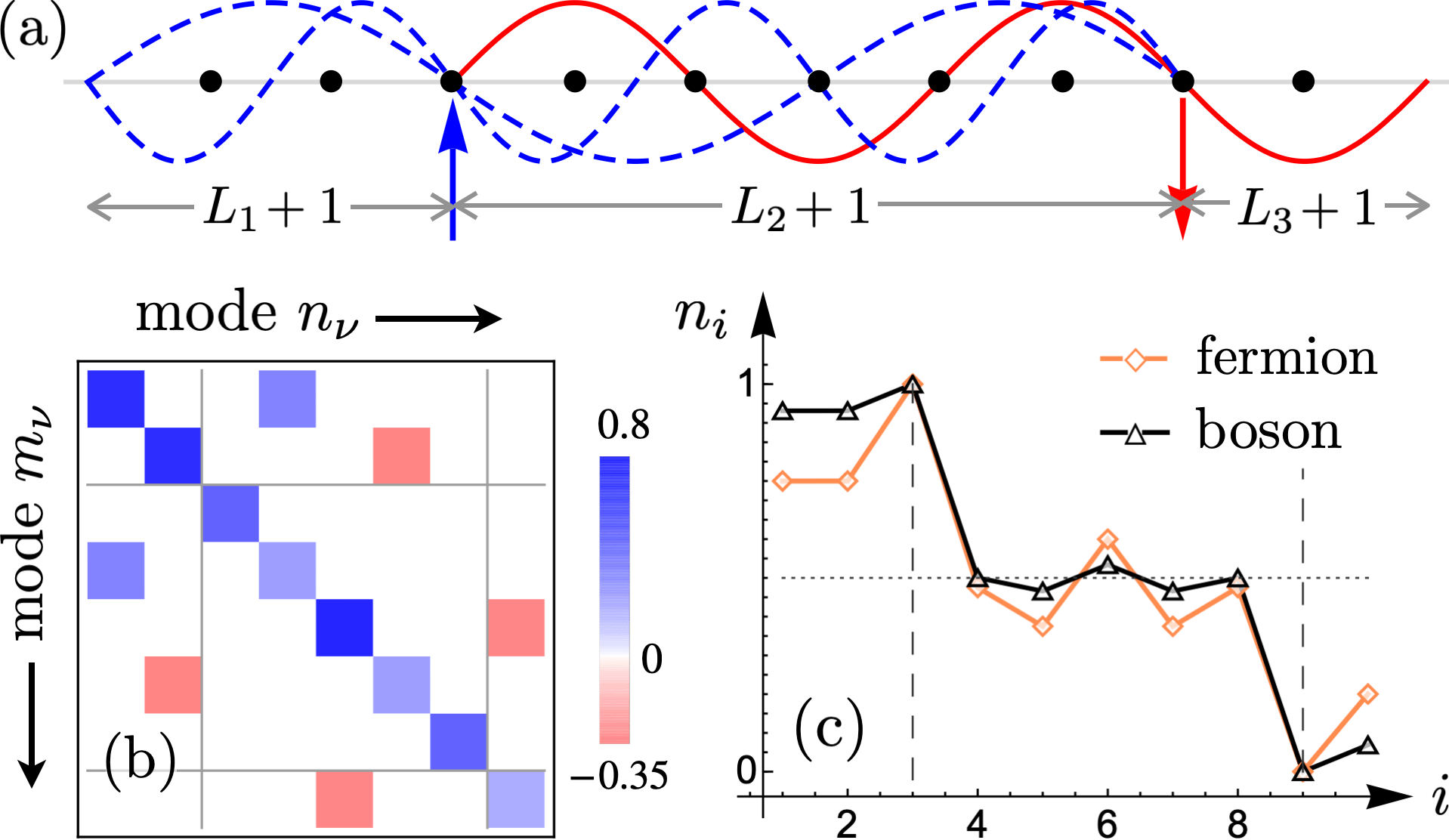}
\caption{\label{zenoplot1}(a) Resonant modes in adjacent segments for a specific pump-loss setup in the Zeno limit, $\gamma_{\pm} \gg J$. (b) Steady-state correlation of free-fermion modes in segments $\nu=1,2,3$, showing coherence among resonant modes. (c) Density of free fermions and hard-core bosons from exact diagonalization.
}
\end{figure}

This picture breaks down if any two modes in neighboring segments are resonant. Then they can remain coherent via the correlated pump and loss, producing characteristic density waves as in Fig.~\ref{zenoplot1}. To understand this feature, note the single-particle modes in segment $\nu$ are given by $\smash{\hat{F}^{(\nu)}_{m_{\nu}} \pull=\pull \sum_{j_{\nu}}\pull c^{(\nu)}_{m_{\nu},j_{\nu}} \hat{f}_{j_{\nu}}}$, with amplitudes $\smash{c^{(\nu)}_{m_{\nu},j_{\nu}}}$ as in Eq.~\eqref{modes}, for $m_{\nu} = 1,\dots,L_{\nu}$, where $\smash{L_{\nu}}$ is the number of sites. They have energies $\smash{\varepsilon^{(\nu)}_{m_{\nu}} = -2\hbar J \cos k^{(\nu)}_{m_{\nu}}}$, where $\smash{k^{(\nu)}_{m_{\nu}} := \pi m_{\nu}/(L_{\nu}\pull+\hspace{-0.04cm} 1)}$. Thus, off-resonant modes dephase at a rate $\Delta\varepsilon/\hbar \sim J$, much faster than the dissipative coupling $\Gamma_{\pm}$. As a result, for $t\gtrsim 1/\Gamma_{\pull \pm}$, adiabatic elimination gives $\langle \hat{F}^{(\nu)\dagger}_{m_{\nu}}\pull \hat{F}^{(\nu+1)}_{m_{\nu+1}}\rangle \pull\approx\pull \Lambda^{(\nu,\nu+1)}_{m_{\nu},m_{\nu+1}} T^{(\nu,\nu+1)}_{m_{\nu},m_{\nu+1}}$, where $\smash{\Lambda^{(\nu,\nu^{\prime})}_{m_{\nu},m_{\nu^{\prime}}}}$ is 1 for resonant modes and 0 otherwise. Since the energies are set by the wavenumber $\smash{k^{(\nu)}_{m_{\nu}}}$, the resonance condition is equivalent to a single plane wave fitting into both segments $\nu$ and $\nu+1$, as in Fig.~\ref{zenoplot1}(a). The corresponding modes are seen to have nonzero steady-state correlation in Fig.~\ref{zenoplot1}(b). Such modes exist iff $L_{\nu}\pull+1$ divides $L_{\nu+1}\pull+\hspace{-0.01cm}1$, or vice versa, making these arrangements special. Within each segment, one finds $\smash{\langle \hat{F}^{(\nu)\dagger}_{m_{\nu}}\pull \hat{F}^{(\nu)}_{n_{\nu}}\rangle \pull\approx\pull \delta_{m_{\nu},n_{\nu}} N^{(\nu)}_{m_{\nu}}}$, as in Sec.~\ref{nonthermal}, where $\smash{N^{(\nu)}_{m_{\nu}}}\hspace{-0.03cm}$ are the mode occupations. These are altered by the resonances, producing density waves in the two coupled segments [Fig.~\ref{zenoplot1}(c)], which are stronger in free fermions than in hard-core bosons.

\begin{figure}
\centering
\includegraphics[width=\columnwidth]{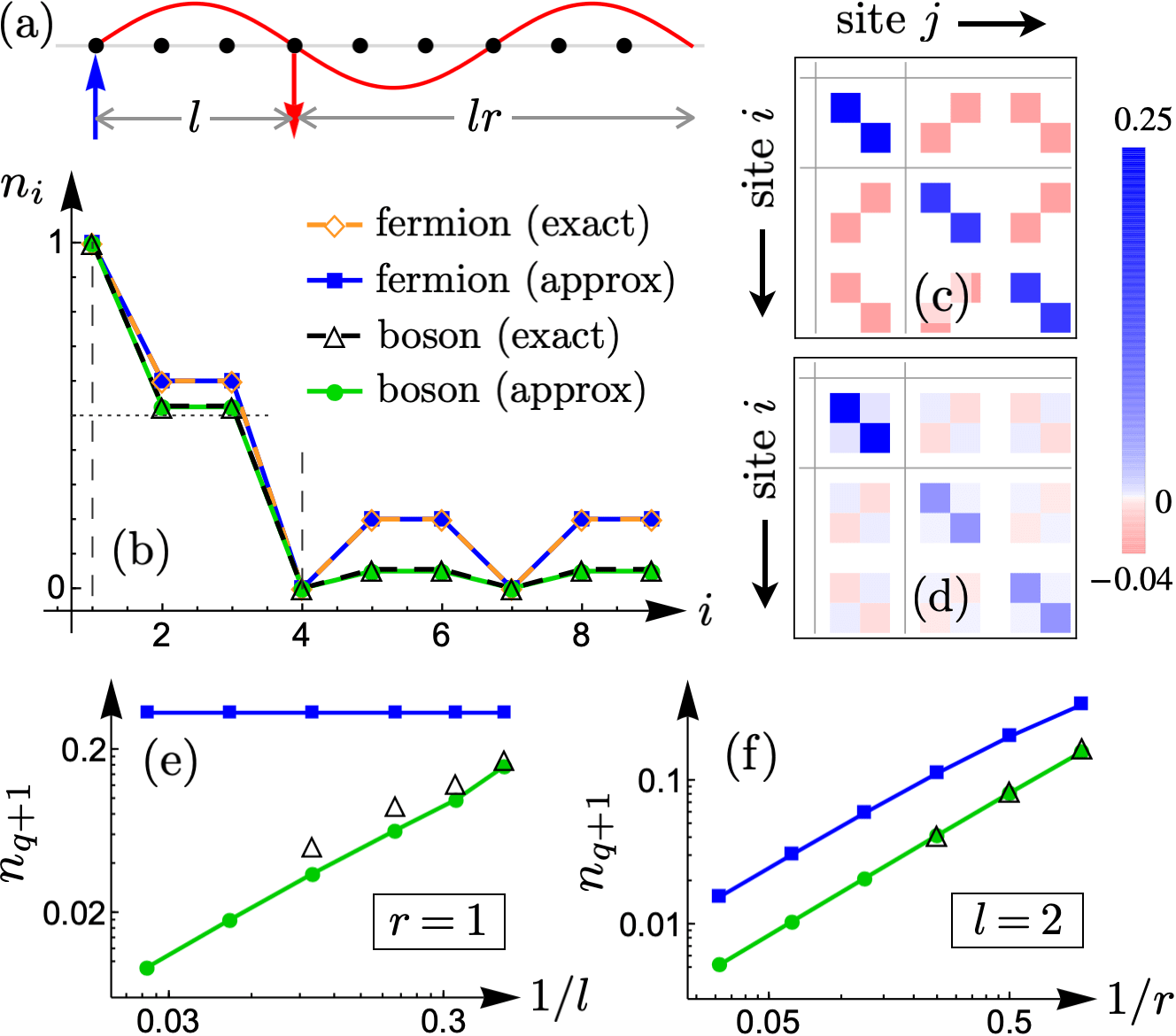}
\caption{\label{zenoplot2}(a) Lowest-energy resonant mode in a class of setups parametrized by integers $l=3$ and $r=2$. (b) Steady-state density of free fermions and hard-core bosons in the Zeno limit from exact diagonalization and our ansatz [Eq.~\eqref{zenoratesfermion}]. (c)--(d) Density correlations of the fermions and bosons, respectively. (e)--(f) Scaling of the density bump $n_{q+1}$ with system size, $q$ being the loss site, with same color convention as in (b). 
}
\end{figure}

These attributes can be explained by approximate rate equations for the modes. For free fermions, one finds (see derivation in \hyperref[srateeqsstrong]{Supplement} \cite{supplement})
\begin{align}
\nonumber \hspace{-0.335cm}\dot{N}^{(3)}_{m_3} \pull &\approx\hspace{-0.02cm} -\Gamma_{\pull -} \hspace{-0.02cm} \Big[|u^{(3)}_{m_3}|^2 N^{(3)}_{m_3} \hspace{-0.03cm}+\hspace{-0.04cm}\text{Re} \hspace{-0.02cm}\sum_{m_2}\hspace{-0.02cm} \Lambda^{(2,3)}_{m_2,m_3} v^{(2)}_{m_2} u^{(3)*}_{m_3} T^{(2,3)}_{m_2,m_3}\hspace{-0.03cm}\Big]\hspace{-0.02cm},\\
&\hspace{-0.63cm}\dot{T}^{(2,3)}_{m_2,m_3} \pull \approx -\Gamma_{\pull f} \hspace{0.03cm} T^{(2,3)}_{m_2,m_3} \pull \hspace{-0.1cm} - \Gamma_{\pull -} v^{(2)*}_{m_2} u^{(3)}_{m_3} \big[\hspace{-0.02cm} N^{(2)}_{m_2} \hspace{-0.07cm}+\hspace{-0.01cm}\pull N^{(3)}_{m_3} \big]/2\hspace{0.05cm},
\label{zenoratesfermion}
\end{align}
where $\smash{u^{(\nu)}_{m_{\nu}}:=c^{(\nu)}_{m_{\nu},1}}$ and $\smash{v^{(\nu)}_{m_{\nu}}:=c^{(\nu)}_{m_{\nu},L_{\nu}}}$ are the mode amplitudes at the boundary where pump or loss occurs, and $\Gamma_{\pull f} \hspace{-0.03cm}:=\hspace{-0.02cm} \frac{\Gamma_{\pull +}}{2} |u^{(2)}_{m_2}|^2 \pull+\pull \frac{\Gamma_{\pull -}}{2} \big[|v^{(2)}_{m_2}|^2 \pull+\pull |u^{(3)}_{m_3}|^2\big]$ is an effective decay rate. The equations for the remaining segments follow by symmetry \cite{supplement}. Note the occupation $\smash{N^{(3)}_{m_3}}$ decays to zero unless it is coupled with a resonant mode $m_2$, producing long-range density-density correlations [Fig.~\ref{zenoplot2}(c)]. The rate equations become exact in the limit $\gamma_{\pm}/J\to\infty$. For hard-core bosons, Eq.~\eqref{zenoratesfermion} gains an interaction term $\dot{T}^{(2,3)}_{m_2,m_3}|_{\text{int}} \pull= 2\push\big\langle\pull\hat{L}^{\text{eff}\dagger}_{-}\hspace{-0.03cm} \hat{F}^{(2)\dagger}_{m_2}\hspace{-0.03cm}  \hat{F}^{(3)}_{m_3}\hspace{-0.03cm}  \hat{L}^{\text{eff}}_{-}\big\rangle$, which can be approximated by pairwise contractions, as in Eq.~\eqref{weakrateboson}. The main result is an increased decay rate, $\smash{\Gamma_{\pull b} = \Gamma_{\pull f} \hspace{-0.03cm}+ 2\push \big\langle \hat{L}^{\text{eff}\dagger}_{-} \hat{L}^{\text{eff}}_{-} \big\rangle}$ (see \hyperref[srateeqsstrong]{Supplement} \cite{supplement}), which weakens the correlations, making the resonant features less prominent.

To further elucidate the characteristic features of hard-core bosons and free fermions, we consider a set of pump-loss configurations with $p=1$, $q=l+1$, and $L=(r+1)l$, for positive integers $l$ and $r$. Here the sink is positioned such that every $r$-th mode in the last segment is resonant with successive modes in the middle [Fig.~\ref{zenoplot2}(a)]. In steady state, this leads to a train of density bumps in the former, with nodes at every $l$-th site [Fig.~\ref{zenoplot2}(b)]. As expected from the preceding paragraph, the bumps are smaller for hard-core bosons, well reproduced by the resonant-modes ansatz. In addition, free fermions show recurrent long-range density correlations of period $2l$ [Fig.~\ref{zenoplot2}(c)], similar to the center-driven case \cite{Dutta2020}. For hard-core bosons, these are less prominent, and cloaked in a uniform background due to interactions [Fig.~\ref{zenoplot2}(d)]. The reduced density fluctuations for bosons originate from a faster decay of correlations. Using $\smash{c^{(2)}_{m_2,j_2} \pull\pull\sim\hspace{-0.03cm} 1/\sqrt{l}}$, $\smash{u^{(3)}_{m_3} \pull\sim\hspace{-0.03cm} 1/\sqrt{lr}}$, and $\smash{N^{(2)}_{m_2}\pull\sim\hspace{-0.03cm} O(1)}$ in Eq.~\eqref{zenoratesfermion} for the resonant modes gives $\smash{\Gamma_{\pull f} \pull\sim \hspace{-0.03cm} O(\Gamma_{\pull\pm}/l)}$ and $\smash{N^{(3)}_{m_3}\pull\sim\hspace{-0.03cm} O(1)}$ for free fermions, and $\Gamma_{\pull b} \pull\sim O(\Gamma_{\pull\pm})$ and $\smash{N^{(3)}_{m_3}\pull\sim O(1/l)}$ for hard-core bosons. Thus, we find a density bump $\smash{n_{q+1}\pull \approx\pull \sum_{m_3}\pull\pull |u^{(3)}_{m_3}|^2 N^{(3)}_{m_3}} \pull\sim\pull 1/L$ in the latter, whereas in the former, $n_{q+1}\pull\sim O(1/r)$, regardless of $l$. This different scaling with system size is evident in Figs.~\ref{zenoplot2}(e) and \ref{zenoplot2}(f). Since the physics is solely determined by geometry, the framework can readily incorporate multiple sources and sinks. Further, Eq.~\eqref{zenoratesfermion} and its bosonic counterpart remain valid in the presence of a trap as long as the spectrum in each segment is nondegenerate.

\section{\label{summary}Summary and outlook}
We have characterized a rich class of steady states that arise in a prototypical setting of two-level systems driven by localized pump and loss. For different arrangements, we find dissipation induced long-range coherence (Fig.~\ref{corrvsdissplot}) and surprising geometric order at strong pump and loss (Fig.~\ref{zenoplot2}). %, that are different for free fermions and hard-core bosons. 
These results could be observed  by measuring local densities and density correlations in state-of-the-art photonic setups \cite{Carusotto2020}. We have developed a general framework to approximate the long-time dynamics at weak and strong dissipation in terms of rate equations for the energy eigenmodes, which would be useful in other systems. Note the strongly dissipative limit is reached in practice for $\gamma_{\pm} \gtrsim 10J$. Quite generally, we find a local source or sink generates correlation between the neighboring sites [Eq.~\eqref{effectiveLs}], which can stabilize long-range order \cite{Dutta2020}. This effect is present whenever the lossy site equilibrates much faster ($\gamma_{\pm}$) than the global relaxation rate ($J/L$) \cite{Popkov2018}, and provides strong motivation for using such dissipation to engineer correlated states of matter \cite{Popkov2020}. Thus, it would be valuable to extend our study to interacting Hamiltonians, such as an XXZ chain \cite{Prosen2010} or a Hubbard model \cite{Kordas2015}, and to higher dimensions \cite{Maghrebi2016} where there is no simple map between hard-core bosons and free fermions \cite{fradkin1989jordan}. It would also be interesting to see how the effect is altered in the continuum. Note the local pump and loss drives current through an interacting medium. Thus, future work could harness such probes to extract useful information about the bulk \cite{Umucalilar2017} and investigate larger questions of nonequilibrium transport \cite{Corman2019}.

\begin{acknowledgments}
%\vspace{0.2cm}
This work was supported by EPSRC Grant No. EP/P009565/1 and by a Simons Investigator Award.
\end{acknowledgments}

\begingroup
\renewcommand{\addcontentsline}[3]{}% Remove functionality of \addcontentsline
\renewcommand{\section}[2]{}% Remove functionality of \section

%%\bibliographystyle{apsrev4-1}
%\bibliography{references}


\endgroup

%\printbibliography
%\begin{thebibliography}{55}
%\end{thebibliography}

\onecolumngrid
\clearpage

\begin{center}
\textbf{\large Supplement for\\``Out-of-equilibrium steady states of a locally driven lossy qubit array"}
\end{center}

\setcounter{equation}{0}
\setcounter{figure}{0}
\setcounter{table}{0}
\setcounter{page}{1}
\setcounter{section}{0}
\setcounter{secnumdepth}{3}
\makeatletter

\renewcommand{\thefigure}{S\arabic{figure}}
\renewcommand{\theequation}{S\arabic{equation}}
\renewcommand{\bibnumfmt}[1]{[S#1]}
\renewcommand{\citenumfont}[1]{S#1}

\renewcommand{\thesection}{\Alph{section}}

\renewcommand{\theHfigure}{S\thefigure}

%\tableofcontents

\vspace{0.4cm}
As described in the main text, we consider hard-core bosons on a 1D lattice modeled by the Hamiltonian
\begin{equation}
\hat{H} = -\hbar J \push\sum\nolimits_{i=1}^{L-1} \hat{b}_{i}^{\dagger} \hat{b}_{i+1} + \hat{b}_{i+1}^{\dagger} \hat{b}_{i}\;,
\label{sbosonhamil}
\end{equation}
where the boson operators satisfy $\smash{[\hat{b}_i,\hat{b}_j]=0}$ and $\smash{[\hat{b}_i,\hat{b}_j^{\dagger}] = (-1)^{\hat{n}_i} \delta_{ij}}$, for site occupation $n_i \in \{0,1\}$. The bosons are coupled to Markovian reservoirs that inject particles at site $p$ with rate $\gamma_{+}$ and removes particles from site $q$ with rate $\gamma_{-}$. The resulting dynamics are modeled by a master equation for the density matrix $\hat{\rho}$
\begin{equation}
\frac{d\hat{\rho}}{dt} = \mathcal{L}\hat{\rho} := -\frac{{\rm i}}{\hbar}\push [\hat{H},\hat{\rho}] \push + \sum\nolimits_{\alpha=\pm}\pull \hat{L}_{\alpha} \hat{\rho}\hat{L}_{\alpha}^{\dagger} - \frac{1}{2} \{\hat{L}_{\alpha}^{\dagger} \hat{L}_{\alpha}, \hat{\rho}\}\;,
\label{smastereqn}
\end{equation}
where $\smash{\hat{L}_+ := \sqrt{\gamma_+}\push\hat{b}_p^{\dagger}}$ and $\smash{\hat{L}_- := \sqrt{\gamma_-}\push\hat{b}_q}$ are Lindblad operators describing the incoherent pump and loss, respectively. The system is equivalent to a spin-1/2 XX chain with local spin flips if one identifies $\hat{b}_i$ with the spin lowering operator. It can also be mapped onto fermions by a Jordan-Wigner transformation,
\begin{equation}
\hat{f}_j = (-1)^{\sum_{i<j} \hat{n}_i} \hat{b}_j\;,
\label{sJordanWigner}
\end{equation}
where $\smash{\{\hat{f}_i,\hat{f}_j\}=0}$ and $\smash{\{\hat{f}_i,\hat{f}_j^{\dagger}\} = \delta_{ij}}$. The transformed Hamiltonian describes free fermions,
\begin{equation}
\hat{H} = -\hbar J \push\sum\nolimits_{i=1}^{L-1} \hat{f}_{i}^{\dagger} \hat{f}_{i+1} + \hat{f}_{i+1}^{\dagger} \hat{f}_{i}\;.
\label{sfermionhamil}
\end{equation}
However, the Lindblad operators $\smash{\hat{L}_{\pm}}$ are nonlocal in the fermions, mediating interactions.

\section{\label{senddrivesol}Closed-form solution for end drives}
As shown in Ref.~\cite{sProsen2008} and discussed in the main text, for pump and loss at opposite ends ($p=1$, $q=L$), the dynamics are identical to those of free fermions, i.e., with $\smash{\hat{L}_+ := \sqrt{\gamma_+}\push\hat{f}_1^{\dagger}}$ and $\smash{\hat{L}_- := \sqrt{\gamma_-}\push\hat{f}_L}$ in Eq.~\eqref{smastereqn}. The steady state is unique \cite{sProsen2012a} and has been solved exactly in terms of a matrix product ansatz \cite{svznidarivc2010matrix}. Here we present a closed-form solution by a more direct approach.%, and extract the single-particle correlations.

We adopt a thermofield representation \cite{sumezawa1995advanced, smedvedyeva2016exact} %of the Liouvillian $\smash{\mathcal{L}}$, 
where one defines a new set of operators $\smash{\tilde{f}_i}$ and $\smash{\tilde{f}_i^{\dagger}}$ that act on the density matrix by right multiplication, i.e., $\smash{\tilde{f}_i \rho := \rho f_i}$ and $\smash{\tilde{f}_i^{\dagger} \rho := \rho f_i^{\dagger}}$. Note we have omitted the hat for operators to reduce clutter. It is straightforward to verify the following relations for any two operators $a$ and $b$:% $[a,\tilde{b}]=0$
\begin{equation}
[a,\tilde{b}]=0 \push ,\;\; (ab)\tilde{\;}=\tilde{b}\tilde{a} \push, \;\;  \{\tilde{a},\tilde{b}\} = \{a,b\}\tilde{\;} \push, \;\text{and}\;\;  [\tilde{a},\tilde{b}] = [b,a]\tilde{\;} \push.
\label{sthermorels}
\end{equation}
The Liouvillian $\mathcal{L}$ in Eq.~\eqref{smastereqn} can be expressed in this notation as
\begin{equation}
\mathcal{L} = {\rm i} J \mathcal{T} + \gamma_{+} \mathcal{D}^+_1 + \gamma_{-} \mathcal{D}^-_L \;,
\label{sliouvillian}
\end{equation}
where
\begin{subequations}
\begin{align}
\mathcal{T} &:= \sum\nolimits_{i=1}^{L-1} f_{i+1}^{\dagger} f_i + \tilde{f}_{i+1}^{\dagger} \tilde{f}_i + f_{i}^{\dagger} f_{i+1} + \tilde{f}_{i}^{\dagger} \tilde{f}_{i+1} \;,
\label{sdefineT}\\
\mathcal{D}^+_i &:= f_i^{\dagger} \tilde{f}_i - \big( f_i f_i^{\dagger} + \tilde{f}_i^{\dagger} \tilde{f}_i \big)/2 \;, 
\label{sdefineDplus}\\
\text{and} \quad \mathcal{D}^-_i &:= f_i \tilde{f}_i^{\dagger} - \big( f_i^{\dagger} f_i + \tilde{f}_i \tilde{f}_i^{\dagger} \big)/2\;.
\label{sdefineDminus}
\end{align}
\end{subequations}
%(\tilde{H} - H)/(\hbar J) =
To write down the steady state, we define the generators
\begin{subequations}
\begin{align}
\mathcal{W}_{i,j} &:= [f_i^{\dagger}, \{f_j, \boldsymbol{\cdot} \}] = \big(f_i^{\dagger} - \tilde{f}_i^{\dagger} \big) \big(f_j + \tilde{f}_j \big)\;,
\label{sdefineW}\\
\text{and}\quad \mathcal{A} &:= \sum\nolimits_{i=1}^{L-1} \mathcal{W}_{i+1,i} - \mathcal{W}_{i,i+1}\;.
\label{sdefineA}
\end{align}
\end{subequations}
%\\\text{and}\quad &\mathcal{B} := \sum\nolimits_{i=2}^{L-1} \mathcal{W}_{i,i}\;.\label{sdefineB}
We will show the steady state is given by $\smash{\rho = e^{\mathcal{G} \tau} \rho_0}$, where %$\rho_0$ is a product state with occupation $\gamma_+/(\gamma_+ \pull+ \gamma_-)$, and
\begin{align}
\mathcal{G} &= \gamma_- \mathcal{W}_{1,1} - \gamma_+ \mathcal{W}_{L,L} + 2\push {\rm i} J \mathcal{A} + (\gamma_- \pull - \gamma_+) \push\textstyle{\sum\nolimits_{j=2}^{L-1}}\push \mathcal{W}_{j,j}\;,
\label{sgenerator}\\
\tau &= (\gamma_+ \pull + \gamma_-)^{-1} \big[1+4J^2/(\gamma_+ \gamma_-)\big]^{-1}\push,
\label{sprefactor}
\end{align}
and $\rho_0$ is a uniform product state with occupation $n_0=\gamma_+/(\gamma_+ \pull + \gamma_-)$,
\begin{equation}
\rho_0 = \prod\nolimits_{i=1}^L n_0 f_i^{\dagger} f_i + (1-n_0) f_i f_i^{\dagger}\;.
\label{sprodstate}
\end{equation}
%\sum\nolimits_{i=2}^{L-1} \mathcal{W}_{i,i}
To prove that $\mathcal{L} \rho = 0$, we first use the identities in Eq.~\eqref{sthermorels} to find the commutators
\begin{subequations}
\begin{align}
[\mathcal{T}, \mathcal{W}_{j,j}] &= (1-\delta_{j,1}) (\mathcal{W}_{j-1,j} - \mathcal{W}_{j,j-1}) + (1-\delta_{j,L}) (\mathcal{W}_{j+1,j} - \mathcal{W}_{j,j+1})\;,
\label{sTW}\\
[\mathcal{T}, \mathcal{A}] &= 2\push (\mathcal{W}_{1,1} - \mathcal{W}_{L,L})\;,
\label{sTA}\\
[\mathcal{D}^{\pm}_i, \mathcal{W}_{j,j}] &= - \delta_{i,j} \mathcal{W}_{i,i}\;,
\label{sDW}\\
[\mathcal{D}^{\pm}_i, \mathcal{A}] &= [(1-\delta_{i,1}) (\mathcal{W}_{i-1,i} - \mathcal{W}_{i,i-1}) - (1-\delta_{i,L}) (\mathcal{W}_{i+1,i} - \mathcal{W}_{i,i+1})]/2\;.
\label{sDA}
\end{align}
\end{subequations}
Using these results in Eqs.~\eqref{sliouvillian} and \eqref{sgenerator} yields
\begin{equation}
[\mathcal{L},\mathcal{G}\tau] = \frac{\gamma_+ \gamma_-}{\gamma_+ \pull + \gamma_-} (\mathcal{W}_{L,L} - \mathcal{W}_{1,1})\;.
\label{sLG}
\end{equation}
Further, one can show $[\mathcal{W}_{i,i},\mathcal{G}]=0$ $\forall i$. Thus, $[[\mathcal{L},\mathcal{G}\tau],\mathcal{G}\tau]=0$ and $\smash{[\mathcal{L},e^{\mathcal{G}\tau}] = e^{\mathcal{G}\tau} [\mathcal{L},\mathcal{G}\tau]}$. It thereby follows that
\begin{equation}
\mathcal{L} \rho = e^{\mathcal{G}\tau} (\mathcal{L} \rho_0 + [\mathcal{L},\mathcal{G}\tau] \rho_0) \;.
\label{sLrho}
\end{equation}
The expression within parentheses can be evaluated by noting that $\rho_0 = (\gamma_+/\gamma_-)^{N}$ up to normalization, where $N$ is the total number operator. Hence, $[H,\rho_0]=0$, or $\mathcal{T} \rho_0 = 0$. One also finds, using Eqs.~\eqref{sprodstate} and \eqref{sLG},
\begin{subequations}
\begin{align}
\gamma_+ \mathcal{D}^+_1 \rho_0 &= \big[(\gamma_+ \pull + \gamma_-) f_1^{\dagger} f_1 - \gamma_+ \big] \rho_0 \;,
\label{sDplusrho0}\\
\gamma_- \mathcal{D}^-_L \rho_0 &= \big[\gamma_+ \pull - (\gamma_+ \pull + \gamma_-) f_L^{\dagger} f_L \big] \rho_0 \;,
\label{sDminusrho0}\\
[\mathcal{L},\mathcal{G}\tau] \rho_0 &= (\gamma_+ \pull + \gamma_-) \big(f_L^{\dagger} f_L - f_1^{\dagger} f_1 \big) \rho_0\;.
\label{scomrho0}
\end{align}
\end{subequations}
Substituting the above results in Eq.~\eqref{sLrho} gives $\mathcal{L}\rho = 0$, thus showing $\rho$ is indeed the steady state.

The computation of $\rho$ can be simplified by noting the generators in Eq.~\eqref{sgenerator} commute with one another, $[\mathcal{W}_{i,i},\mathcal{A}] = 0$ and $[\mathcal{W}_{i,i},\mathcal{W}_{j,j}]=0$ $\forall \push i,j$. Thus, $e^{\mathcal{G}\tau}$ factorizes into a product of exponentials. Further, $\mathcal{W}_{i,i}$ acts locally and one can show $\mathcal{W}_{i,i}^2 = 0$ $\forall i$, which means the action of these local generators on $\rho_0$ can be written explicitly, yielding
\begin{equation}
\rho = e^{2\push {\rm i} J \tau \mathcal{A}} \push \prod\nolimits_{i=1}^L (n_0 \pull+\pull\Delta_i) f_i^{\dagger} f_i + (1 \pull-\pull n_0 \pull-\pull \Delta_i) f_i f_i^{\dagger}\;,
\label{srhosimple}
\end{equation}
where $\Delta_i := [(1 \pull-\pull \delta_{i,L}) \gamma_- \pull- (1 \pull-\pull \delta_{i,1}) \gamma_+]\push\tau$. Furthermore, $\mathcal{A}^{L+1}=0$, so the above exponential reduces to a sum of the first $L+1$ terms in its power series, which can be computed iteratively using $\smash{\mathcal{A} \push \sigma = \sum_{i=1}^{L-1} [f_{i+1}^{\dagger},\{f_i,\sigma\}] - \text{H.c.}}$. Such a simple algebraic structure of the steady state is related to the $q$-deformed SU(2) symmetry of the XXZ chain \cite{skarevski2013exact}.

For weak dissipation ($\gamma_{\pm} \ll J$), $J\tau \approx \gamma_{+} \gamma_-/[4J(\gamma_+ \pull + \gamma_-)] + O((\gamma_{\pm}/J)^3)$ and $\Delta_i \sim O((\gamma_{\pm}/J)^2)$. Thus, to linear order in $\gamma_{\pm}/J$, Eq.~\eqref{srhosimple} gives the steady state
\begin{equation}
\rho \approx \rho_0 + {\rm i} \push \frac{\gamma_+ \gamma_-}{2J(\gamma_+ \pull + \gamma_-)} \push \mathcal{A} \rho_0 = \rho_0 + \Big( {\rm i} \push \frac{\gamma_+ \pull + \gamma_-}{2J} \sum\nolimits_{j=1}^{L-1} f_{j+1}^{\dagger} f_j \push \rho_0 + \text{H.c.} \Big) \push,
\label{sendweak}
\end{equation}
where we have used Eq.~\eqref{sprodstate} to simplify $\mathcal{A}\rho_0$. The perturbation to $\rho_0$ induces nearest-neighbor correlations,
\begin{equation}
\langle b_j^{\dagger} b_{j+1} \rangle = \langle f_j^{\dagger} f_{j+1} \rangle \approx {\rm i} \push \frac{\gamma_+ \gamma_-}{2 J (\gamma_+ \pull + \gamma_-)} \;.
\label{sendcorrweak}
\end{equation}
For strong dissipation ($\gamma_{\pm} \gg J$), $J\tau \approx J/(\gamma_+ \pull + \gamma_-) + O((J/\gamma_{\pm})^3)$, and the product state in Eq.~\eqref{srhosimple} approaches $\rho_Z$ where the first site is filled, the last site is empty, and the other sites are in a product state with occupation $1-n_0$,
\begin{equation}
\rho_Z := f_1^{\dagger} f_1 f_L f_L^{\dagger} \push \prod\nolimits_{i=2}^{L-1} (1-n_0) f_i^{\dagger} f_i + n_0 f_i f_i^{\dagger}\push.
\label{srhoz}
\end{equation}
To first order in $J/\gamma_{\pm}$, the steady state is given by
\begin{equation}
\rho \approx \rho_Z + \bigg( {\rm i} \push \frac{2J}{\gamma_+ \pull + \gamma_-} \sum\nolimits_{j=1}^{L-1} [f_{j+1}^{\dagger}, \{f_j, \rho_Z \}] + \text{H.c.} \pull \bigg) \push,
\label{sendstrong}
\end{equation}
which again yields nearest-neighbor correlations,
\begin{equation}
\langle b_j^{\dagger} b_{j+1} \rangle = \langle f_j^{\dagger} f_{j+1} \rangle \approx {\rm i} \push \frac{2J}{\gamma_+ \pull + \gamma_-} \;.
\label{sendcorrstrong}
\end{equation}
Hence, the correlations are limited to nearest neighbors in both limits, and are purely imaginary, which corresponds to a probability current from the source at the first site to the sink at the last site.

%\begin{equation}
%\mathcal{G} = \bigg(\pull 1+\frac{4J^2}{\gamma_+ \gamma_-}\bigg)^{\pull\pull -1} \bigg(\frac{\gamma_-}{\gamma_+ \pull+ \gamma_-} \mathcal{W}_{1,1} - \frac{\gamma_+}{\gamma_+ \pull+ \gamma_-} \mathcal{W}_{L,L} + \frac{2\push {\rm i} J}{\gamma_+ \pull+ \gamma_-} \mathcal{A} + \frac{\gamma_- \pull -\gamma_+}{\gamma_+ \pull+ \gamma_-} \sum_{i=2}^{L-1} \mathcal{W}_{i,i} \bigg)\;.
%\label{sgenerator}
%\end{equation}
%Using the identities in Eq.~\eqref{sthermorels}, one can show $[\mathcal{W}_{i,i},\mathcal{W}_{j,j}]=0$ and $[\mathcal{W}_{i,i},\mathcal{A}] = 0$ $\forall \push i.j$. Also, the action of $\mathcal{A}$ on a density matrix can be computed as

\section{\label{sdipolesol}Perturbative solutions for a dipole drive}
In Sec.~\ref{corrvsrate} of the main text, we described a ``dipole'' arrangement of the pump and loss, where increasing dissipation establishes long-range coherence, in sharp contrast to the end-driven case studied above. In this ``dipole'' setup, the pump and loss occur at neighboring sites in the middle, $p=L/2$ and $q=L/2+1$ for even $L$. Here we derive the perturbation results for weak and strong dissipation quoted in the main text.

The Liouvillian $\mathcal{L}$ in Eq.~\eqref{smastereqn} can be restated as
\begin{equation}
\mathcal{L} = -({\rm i}/\hbar) [\hat{H}, \boldsymbol{\cdot}\;] + \gamma_+ \mathcal{D}[\hat{b}_{L/2}^{\dagger}] + \gamma_- \mathcal{D}[\hat{b}_{L/2+1}]\;,
\label{sLdipole}
\end{equation}
where $\smash{\mathcal{D}[\hat{x}] \hat{\rho} := \hat{x} \hat{\rho} \hat{x}^{\dagger} - \{\hat{x}^{\dagger} \hat{x}, \hat{\rho}\}/2}$. The steady state at weak dissipation approaches the product state $\smash{\hat{\rho}_0 \propto (\gamma_+/\gamma_-)^{\hat{N}}}$, as in the end-driven geometry [see Eq.~\eqref{sprodstate}]. This is the zeroth order solution, which commutes with the Hamiltonian. %(Recall, $\smash{\hat{N}}$ is the total number operator.) 
The solution to first order in $\gamma_{\pm}/J$ is given by $\hat{\rho}_{\text{w}} \approx \hat{\rho}_0 + \hat{\rho}_1$, such that
\begin{equation}
-({\rm i}/\hbar) [\hat{H}, \hat{\rho}_1] + \gamma_+ \mathcal{D}[\hat{b}_{L/2}^{\dagger}] \hat{\rho}_0 + \gamma_- \mathcal{D}[\hat{b}_{L/2+1}] \hat{\rho}_0 = 0\;.
\label{sdipoleweakpertcond}
\end{equation}
We will show this is satisfied by
\begin{align}
\hat{\rho}_1 =& \hspace{0.12cm} {\rm i}\push \frac{\gamma_+ \pull + \gamma_-}{2J} \big(\hat{Q} - \hat{Q}^{\dagger}\big) \hat{\rho}_0\;,
\label{sdipoleweak}\\
\text{where}\quad \hat{Q} := & \push\sum\nolimits_{k=1}^{L/2} \hat{f}_{L+1-k}^{\dagger} \hat{f}_k\;.
\label{sdefineQ}
\end{align}
First, we calculate the action of the dissipators on the unperturbed solution. As in Eqs.~\eqref{sDplusrho0} and \eqref{sDminusrho0}, we find
\begin{equation}
\gamma_+ \mathcal{D}[\hat{b}_{L/2}^{\dagger}] \hat{\rho}_0 + \gamma_- \mathcal{D}[\hat{b}_{L/2+1}] \hat{\rho}_0 = (\gamma_+ \pull + \gamma_-) (\hat{n}_{L/2} - \hat{n}_{L/2+1}) \hat{\rho}_0\;.
\label{sdipoleweakDrho1}
\end{equation}
Next, we find the commutator $\smash{[\hat{H},\hat{\rho}_1]}$ using Eq.~\eqref{sfermionhamil} and the identity $[ab,cd] = a\{b,c\}d + ca\{b,d\} - \{a,c\}bd - c\{a,d\}b$,
\begin{equation}
-({\rm i}/\hbar) [\hat{H}, \hat{\rho}_1] = -\frac{\gamma_+ \pull + \gamma_-}{2} \Big(\sum\nolimits_{i=1}^{L-1} [\hat{f}_i^{\dagger} \hat{f}_{i+1} + \hat{f}_{i+1}^{\dagger} \hat{f}_{i},\hat{Q}] + \text{H.c.}\Big) \hat{\rho}_0 = (\gamma_+ \pull + \gamma_-) (\hat{n}_{L/2+1} - \hat{n}_{L/2}) \hat{\rho}_0\;.
\label{sdipoleweakHrho0}
\end{equation}
Combining Eqs.~\eqref{sdipoleweakDrho1} and \eqref{sdipoleweakHrho0} readily gives the first-order condition in Eq.~\eqref{sdipoleweakpertcond}. Note the perturbation $\hat{\rho}_1$ induces coherence between reflection-symmetric sites $k$ and $L+1-k$, leading to the single-particle correlations (for $k \leq L/2)$
\begin{align}
\nonumber \langle \hat{b}_k^{\dagger} \hat{b}_{L+1-k} \rangle_{\text{w}} &\approx {\rm i}\push \frac{\gamma_+ \pull + \gamma_-}{2J} \push \text{Tr} \big( \hat{b}_k^{\dagger} \hat{b}_{L+1-k} \push \hat{f}_{L+1-k}^{\dagger} \hat{f}_{k} \push \hat{\rho}_0 \big) \\
\nonumber &= {\rm i}\push \frac{\gamma_+ \pull + \gamma_-}{2J} \push \text{Tr} \Big[ \hat{b}_k^{\dagger} \hat{b}_{L+1-k} \push \hat{b}_{L+1-k}^{\dagger} \hat{b}_{k} \prod\nolimits_{i=k+1}^{L-k} (-1)^{\hat{n}_i} \hat{\rho}_0 \Big] \hspace{1cm} [\text{using Eq.~\eqref{sJordanWigner}}]\\
\nonumber &= {\rm i}\push \frac{\gamma_+ \pull + \gamma_-}{2J} \push \text{Tr} \Big[ \hat{n}_k (1-\hat{n}_{L+1-k}) \prod\nolimits_{i=k+1}^{L-k} (-1)^{\hat{n}_i} \hat{\rho}_0 \Big] \\
\nonumber &= {\rm i}\push \frac{\gamma_+ \pull + \gamma_-}{2J} \frac{\gamma_+}{\gamma_+ \pull + \gamma_-} \frac{\gamma_-}{\gamma_+ \pull + \gamma_-} \bigg(\frac{\gamma_- \pull - \gamma_+}{\gamma_+ \pull + \gamma_-}\bigg)^{\hspace{-0.1cm} L-2k} \hspace{2.1cm} [\text{using Eq.~\eqref{sprodstate}}] \\
&= {\rm i}\push \frac{\gamma_+ \gamma_-}{2J (\gamma_+ \pull + \gamma_-)} \bigg(\frac{\gamma_+ \pull - \gamma_-}{\gamma_+ \pull + \gamma_-}\bigg)^{\hspace{-0.1cm} L-2k} .
\label{sdipoleweakcorr}
\end{align}
Thus, the correlations fall off exponentially with distance due to the string, and vanish for $k<L/2$ if $\gamma_+ = \gamma_-$.

At strong dissipation ($\gamma_{\pm} \gg J$), the steady state approaches a step where all sites $i \leq L/2$ are filled and all sites $i>L/2$ are empty (see Sec.~\ref{resonance} in the main text for a physical explanation). This pure state is expressed as
\begin{equation}
\hat{\rho}_{\text{step}} = \hat{n}_1 \dots \hat{n}_{L/2}\push (1-\hat{n}_{L/2+1}) \dots (1-\hat{n}_L) \;,
\label{sdipolestep}
\end{equation}
and annihilated by the dissipators $\smash{\mathcal{D}[\hat{b}_{L/2}^{\dagger}]}$ and $\smash{\mathcal{D}[\hat{b}_{L/2+1}]}$ in Eq.~\eqref{sLdipole}. To first order in $J/\gamma_{\pm}$, the steady state is of the form $\hat{\rho}_{\text{s}} \approx \hat{\rho}_{\text{step}} + \hat{\rho}_1$, such that
\begin{equation}
-({\rm i}/\hbar) [\hat{H}, \hat{\rho}_{\text{step}}] + \gamma_+ \mathcal{D}[\hat{b}_{L/2}^{\dagger}] \hat{\rho}_1 + \gamma_- \mathcal{D}[\hat{b}_{L/2+1}] \hat{\rho}_1 = 0\;.
\label{sdipolestrongpertcond}
\end{equation}
We will show the perturbation $\hat{\rho}_1$ is again generated by the operator $\hat{Q}$ in Eq.~\eqref{sdefineQ},
\begin{equation}
\hat{\rho}_1 =  {\rm i} \push \frac{2 J}{\gamma_+ \pull + \gamma_-} \big(\hat{Q}\push\hat{\rho}_{\text{step}} - \text{H.c.}\big)\;.
\label{sdipolestrong}
\end{equation}
First, using the expression for $\smash{\hat{H}}$ in Eq.~\eqref{sfermionhamil}, one finds
\begin{align}
\nonumber -({\rm i}/\hbar) [\hat{H}, \hat{\rho}_{\text{step}}] &= {\rm i} \push J \push \big[ \hat{f}_{L/2+1}^{\dagger} \hat{f}_{L/2}, \hat{n}_1 \dots \hat{n}_{L/2}\push (1-\hat{n}_{L/2+1}) \dots (1-\hat{n}_L) \big] +\text{H.c.} \\
&= {\rm i} \push J \hat{f}_{L/2+1}^{\dagger} \hat{f}_{L/2} \push \hat{\rho}_{\text{step}} + \text{H.c.}\;.
\label{sdipoleHrhostep}
\end{align}
Next, acting the dissipator $\smash{\mathcal{D}[\hat{b}_{L/2}^{\dagger}]}$ on $\hat{\rho}_1$ and using $\smash{\hat{\rho}_{\text{step}}\hat{b}_{L/2} =  0}$ yields
\begin{equation}
\mathcal{D}[\hat{b}_{L/2}^{\dagger}] \hat{\rho}_1 = - {\rm i} \push \frac{J}{\gamma_+ \pull + \gamma_-} \hat{f}_{L/2} \hat{f}_{L/2}^{\dagger} \hat{Q}\push \hat{\rho}_{\text{step}} +\text{H.c.} = - {\rm i} \push \frac{J}{\gamma_+ \pull + \gamma_-} \hat{f}_{L/2+1}^{\dagger} \hat{f}_{L/2} \push \hat{\rho}_{\text{step}} +\text{H.c.}\;.
\end{equation}
The same result is found for $\smash{\mathcal{D}[\hat{b}_{L/2+1}] \hat{\rho}_1}$. Thus,
\begin{equation}
\gamma_+ \mathcal{D}[\hat{b}_{L/2}^{\dagger}] \hat{\rho}_1 + \gamma_- \mathcal{D}[\hat{b}_{L/2+1}] \hat{\rho}_1 = - {\rm i} \push J \hat{f}_{L/2+1}^{\dagger} \hat{f}_{L/2} \push \hat{\rho}_{\text{step}} + \text{H.c.}\;.
\label{sdipoleDrho1}
\end{equation}
This exactly cancels $\smash{-({\rm i}/\hbar) [\hat{H}, \hat{\rho}_{\text{step}}]}$ in Eq.~\eqref{sdipoleHrhostep}, satisfying the first-order condition in Eq.~\eqref{sdipolestrongpertcond}. This single-particle correlations can be calculated by a similar procedure as in Eq.~\eqref{sdipoleweakcorr}, yielding (for $k \leq L/2)$
\begin{align}
\nonumber \langle \hat{b}_k^{\dagger} \hat{b}_{L+1-k} \rangle_{\text{s}}
&\approx {\rm i}\push \frac{2J}{\gamma_+ \pull + \gamma_-} \push \text{Tr} \Big[ \hat{n}_k (1-\hat{n}_{L+1-k}) \prod\nolimits_{i=k+1}^{L-k} (-1)^{\hat{n}_i} \hat{\rho}_{\text{step}} \Big] \\
&= {\rm i}\push \frac{2J}{\gamma_+ \pull + \gamma_-} (-1)^{L/2-k}\;.
\label{sdipolestrongcorr}
\end{align}
Now the string gives rise to constant-amplitude oscillations, stabilizing long-range coherence. Note the steady states in Eqs.~\eqref{sdipoleweak} and \eqref{sdipolestrong} can be written as a compact matrix product operator, as in Refs.~\cite{svznidarivc2010matrix, skarevski2013exact}.

\section{\label{sotherdipoles}Geometric resonance and long-range coherence in dipole setups}
\begin{figure}[h]
\vspace{0.3cm}
\centering
\includegraphics[width=\textwidth]{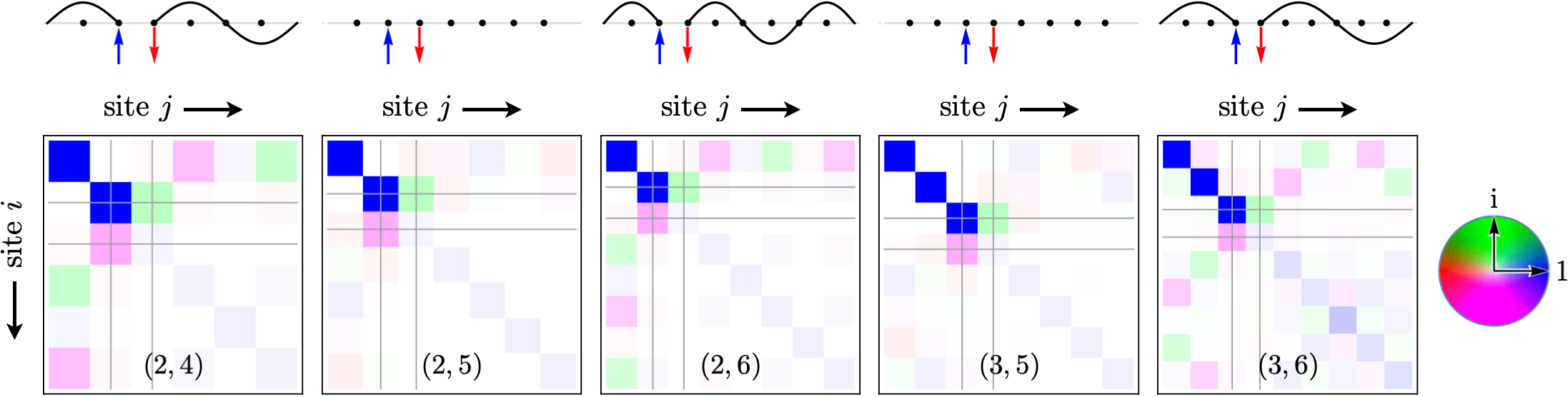}
\caption{\label{otherdipolefig}Top panel shows schematic ``dipole'' setups where the pump and loss divide the system into two segments which may share a resonant mode, shown by black lines. Bottom panel shows steady-state correlations $\smash{\langle\hat{b}_i^{\dagger}\hat{b}_j\rangle}$ for the corresponding setups at strong dissipation, $\smash{\gamma_+\pull=\gamma_-=20J}$. The numbers in parentheses give the size of the segments, $p$ and $L-p$. The first four plots are obtained from exact diagonalization and the last plot is obtained by averaging over quantum trajectories \cite{sDaley2014}.
}
\end{figure}

In the main text, we mentioned that the dissipation induced long-range coherence found above can be generalized to other ``dipole'' setups where the pump and loss act on neighboring sites ($q=p+1$), provided $p$ divides $L-p$, or vice versa. Here we present numerical examples. Figure~\ref{otherdipolefig} shows the single-particle density matrices in steady state for different dipole arrangements at strong dissipation. Here the pump and loss divide the system into weakly coupled segments, sites 1 through $p$ and sites $q$ through $L$, as explained in Sec.~\ref{resonance} of the main text. Long-range coherence is found when the two segments share a resonant mode, as sketched in the upper panels of Fig.~\ref{otherdipolefig}. This is a higher-order analog of the geometric resonances discussed in the main text. In all of the examples, the coherences are limited to nearest neighbors at weak dissipation ($\gamma_{\pm} \ll J$). Hence, the resonant geometries constitute a family of setups where long-range coherence is stabilized by increasing the pump and loss rates.

\section{\label{srateeqsweak}Rate equations for weak dissipation}
In Sec.~\ref{nonthermal} of the main text, we explained that when the dissipation is weak compared to the energy splitting of the single-particle modes, the modes become uncorrelated, leading to simplified rate equations for the mode occupations which characterize the steady state. Here we derive these approximate rate equations.

First, we consider the equation of motion for the expectation of a general observable $\hat{A}$. Substituting $\langle \hat{A} \rangle = \text{Tr}(\hat{A} \hat{\rho} )$ into Eq.~\eqref{smastereqn} and using the cyclic property of trace, one finds
\begin{equation}
\frac{d \langle \hat{A} \rangle}{dt} = \text{Tr} \left(\pull\hat{A} \push\frac{d \hat{\rho}}{dt} \right)
= \frac{{\rm i}}{\hbar} \langle [\hat{H}, \hat{A}] \rangle - \text{Re} \push \sum\nolimits_{\alpha=\pm} \langle \hat{L}_{\alpha}^{\dagger} [\hat{L}_{\alpha},\hat{A}] \rangle\;.
\label{sobseom}
\end{equation}
The single-particle modes of the Hamiltonian in Eq.~\eqref{sfermionhamil} are of the form $\smash{\hat{F}_m = \sum_{j} c_{m,j} \hat{f}_j}$, such that $\sum_{m} c_{m,i}^* c_{m,j} = \delta_{ij}$. One can invert this relation to find
\begin{equation}
\hat{f}_i = \sum\nolimits_{m=1}^L c_{m,i}^{*} \hat{F}_m \push .
\label{sFtof}
\end{equation}
The number operator for a mode is given by $\smash{\hat{N}_m := \hat{F}_m^{\dagger} \hat{F}_m}$ which, by definition, commutes with $\hat{H}$. The equation of motion for the mode occupation $\smash{N_m := \langle \hat{N}_m \rangle}$ is simpler if one has pump and loss of free fermions, i.e., $\smash{\hat{L}_{+} = \sqrt{\gamma_+}\push \hat{f}_p^{\dagger}}$ and $\smash{\hat{L}_{-} = \sqrt{\gamma_-}\push \hat{f}_q}$. Then Eq.~\eqref{sobseom} yields
\begin{equation}
\dot{N}_m = -\gamma_+ \push \text{Re} \push \langle \hat{f}_p [\hat{f}_p^{\dagger}, \hat{N}_m] \rangle - \gamma_- \push \text{Re} \push \langle \hat{f}_q^{\dagger} [\hat{f}_q, \hat{N}_m] \rangle \;.
\label{snm1}
\end{equation}
The commutators can be found using Eq.~\eqref{sFtof} and the relations $\{\hat{F}_m,\hat{F}_n\}=0$ and $\{\hat{F}_m^{\dagger},\hat{F}_n\}=\delta_{m,n}$, which gives
\begin{equation}
\langle \hat{f}_q^{\dagger} [\hat{f}_q, \hat{N}_m] \rangle = c_{m,q}^* \sum\nolimits_n c_{n,q} \langle\hat{F}_n^{\dagger} \hat{F}_m\rangle\;.
\label{sfqdagfqnm}
\end{equation}
Now we approximate the modes to be uncorrelated, i.e., $\smash{\langle\hat{F}_n^{\dagger} \hat{F}_m\rangle \approx \delta_{m,n} N_m}$, obtaining $\smash{\langle \hat{f}_q^{\dagger} [\hat{f}_q, \hat{N}_m] \rangle \approx |c_{m,q}|^2 N_m}$. The pump term in Eq.~\eqref{snm1} is found by exchanging $p \leftrightarrow q$ and particles with holes, yielding
\begin{equation}
\dot{N}_m \approx \gamma_+ |c_{m,p}|^2 \bar{N}_m - \gamma_- |c_{m,q}|^2 N_m\;,
\label{snmfermion}
\end{equation}
where $\smash{\bar{N}_m := 1 - N_m}$ is the hole occupation. %Thus, the incoming and outgoing rates are set by the weights at the pump and loss sites, respectively.

For pump and loss of hard-core bosons, i.e., $\smash{\hat{L}_{+} = \sqrt{\gamma_+}\push \hat{b}_p^{\dagger}}$ and $\smash{\hat{L}_{-} = \sqrt{\gamma_-}\push \hat{b}_q}$, Eq.~\eqref{snm1} is modified as
\begin{equation}
\dot{N}_m = -\gamma_+ \push \text{Re} \push \langle \hat{b}_p [\hat{b}_p^{\dagger}, \hat{N}_m] \rangle - \gamma_- \push \text{Re} \push \langle \hat{b}_q^{\dagger} [\hat{b}_q, \hat{N}_m] \rangle \;.
\label{snmboson1}
\end{equation}
Mapping the bosons onto fermions through Eq.~\eqref{sJordanWigner} and using the expansion in Eq.~\eqref{sFtof}, we find
\begin{equation}
\langle \hat{b}_q^{\dagger} [\hat{b}_q, \hat{N}_m] \rangle 
= c_{m,q}^* \sum\nolimits_n \pull c_{n,q} \langle\hat{F}_n^{\dagger} \hat{F}_m\rangle 
+ \sum\nolimits_{n,n^{\prime}}\pull c_{n,q} c_{n^{\prime},q}^* \push 
\big\langle \hat{F}_n^{\dagger} \push (-1)^{\hat{\mathcal{N}}_q} \big[ (-1)^{\hat{\mathcal{N}}_q}, \hat{N}_m \big] \hat{F}_{n^{\prime}} \big\rangle \;,
\label{sbqdagbqnm}
\end{equation}
where $\smash{\hat{\mathcal{N}}_q := \sum_{i<q}\hat{n}_i}$. The interaction can be simplified by writing $\smash{\hat{N}_m = \sum_{i,j} c_{m,i}^* c_{m,j} \hat{f}_i^{\dagger} \hat{f}_j}$ and noting that $\smash{(-1)^{\hat{\mathcal{N}}_q}}$ transforms $\hat{f}_i$ to $-\hat{f}_i$ only if $i<q$. Thus,
\begin{equation}
(-1)^{\hat{\mathcal{N}}_q} \hat{N}_m (-1)^{\hat{\mathcal{N}}_q} = \sum\nolimits_{i,j} \sigma_{i-q} \sigma_{j-q} c_{m,i}^* c_{m,j} \hat{f}_i^{\dagger} \hat{f}_j\;,
\label{sintsimplify}
\end{equation}
where $\sigma_k = 1$ for $k \geq 0$ and $-1$ for $k<0$. Rewriting the $\smash{\hat{f}_j}$'s in terms of the modes in Eq.~\eqref{sFtof}, one finds
\begin{align}
(-1)^{\hat{\mathcal{N}}_q} \hat{N}_m (-1)^{\hat{\mathcal{N}}_q} 
=& \sum\nolimits_{r,s} \alpha^{(q)}_{m,r} \alpha^{(q)}_{s,m} \hat{F}_r^{\dagger} \hat{F}_s\;,
\label{snqstring}\\
\text{where} \quad \alpha^{(i)}_{m,n} :=& \sum\nolimits_j \sigma_{j-i} c_{m,j}^* c_{n,j}\;.
\label{sdefinealpha}
\end{align}
Substituting Eq.~\eqref{snqstring} into Eq.~\eqref{sbqdagbqnm} gives
\begin{equation}
\langle \hat{b}_q^{\dagger} [\hat{b}_q, \hat{N}_m] \rangle
= c_{m,q}^* \sum\nolimits_n \pull c_{n,q} \langle\hat{F}_n^{\dagger} \hat{F}_m\rangle
+ \sum\nolimits_{n,n^{\prime}}\pull c_{n,q} c_{n^{\prime},q}^* \Big[ \langle \hat{F}_n^{\dagger} \hat{F}_m^{\dagger} \hat{F}_m \hat{F}_{n^{\prime}} \rangle 
- \sum\nolimits_{r,s} \pull \alpha^{(q)}_{m,r} \alpha^{(q)}_{s,m} \langle \hat{F}_n^{\dagger} \hat{F}_r^{\dagger} \hat{F}_s \hat{F}_{n^{\prime}} \rangle \Big]\;.
\label{sbqdagbqnmexpanded}
\end{equation}
Thus, the string gives rise to quartic coupling among the modes. Note the above result is exact for hard-core bosons. We approximate the quartic terms using the product-of-modes ansatz, $\smash{\hat{\rho} \approx N_m |1_m\rangle\langle 1_m| + \bar{N}_m \rangle |0_m\rangle\langle 0_m|}$, presented in the main text (recall, $\bar{x}:=1-x$), finding
\begin{equation}
\langle \hat{F}_m^{\dagger} \hat{F}_{m^{\prime}}^{\dagger} \hat{F}_n \hat{F}_{n^{\prime}} \rangle 
\approx N_m N_{m^{\prime}} (\delta_{m,n^{\prime}} \delta_{m^{\prime},n} - \delta_{m,n} \delta_{m^{\prime},n^{\prime}}) \;,
\label{squarticapprox}
\end{equation}
along with $\smash{\langle\hat{F}_n^{\dagger} \hat{F}_m\rangle \approx \delta_{m,n} N_m}$. Using these expressions in Eq.~\eqref{sbqdagbqnmexpanded} yields
\begin{equation}
\langle \hat{b}_q^{\dagger} [\hat{b}_q, \hat{N}_m] \rangle
\approx N_m \big( \bar{N}_m |c_{m,q}|^2 \hspace{-0.04cm}+\hspace{-0.01cm} n_q\big)\pull - n_q \beta^{(q)}_m\pull + |\kappa^{(q)}_m|^2\;,
\label{sbqdagbqnmfinal}
\end{equation}
where
\begin{equation}
\beta^{(i)}_m :=\sum\nolimits_n\pull N_n |\alpha^{(i)}_{m,n}|^2 \quad \text{and}\quad
\kappa^{(i)}_m := \sum\nolimits_n\pull N_n \alpha^{(i)}_{n,m} c_{n,i}\;.
\label{sdefinebetakappa}
\end{equation}
Note that $n_q$ in Eq.~\eqref{sbqdagbqnmfinal} is the occupation at the loss site, and not a mode index. It is related to the mode occupations as $\smash{n_q := \langle \hat{f}_q^{\dagger} \hat{f}_q \rangle \approx \sum_n N_n |c_{n,q}|^2}$. The analog of Eq.~\eqref{sbqdagbqnmfinal} for the pump term is again found by swapping the particle and hole occupations and exchanging $p \leftrightarrow q$, which gives
\begin{equation}
\langle \hat{b}_p [\hat{b}_p^{\dagger}, \hat{N}_m] \rangle
\approx -\bar{N}_m\hspace{-0.02cm} \big(N_m |c_{m,p}|^2 \hspace{-0.04cm}+\hspace{-0.01cm} \bar{n}_p\big)\pull + \bar{n}_p \bar{\beta}^{(p)}_m\pull - |\kappa^{(p)}_m\pull -\hspace{-0.01cm}c_{m,p}|^2\;.
\label{sbpbpdagnmfinal}
\end{equation}
For the last two terms, we have used $\smash{\sum_n |\alpha^{(i)}_{m,n}|^2 = 1}$ and $\sum_n\alpha^{(i)}_{n,m} c_{n,i} = c_{m,i}$ in Eq.~\eqref{sdefinebetakappa}. Substituting Eqs.~\eqref{sbqdagbqnmfinal} and \eqref{sbpbpdagnmfinal} into Eq.~\eqref{snmboson1} yields the coupled nonlinear rate equations for hard-core bosons,
\begin{equation}
\dot{N}_m \approx \gamma_+\big[\hspace{0.02cm}\bar{N}_m\hspace{-0.02cm} \big(N_m |c_{m,p}|^2 \hspace{-0.04cm}+\hspace{-0.01cm} \bar{n}_p\big)\pull - \bar{n}_p \bar{\beta}^{(p)}_m\pull + |\kappa^{(p)}_m\pull -\hspace{-0.01cm}c_{m,p}|^2 \hspace{0.02cm}\big]
- \gamma_- \big[\hspace{0.02cm} N_m \big( \bar{N}_m |c_{m,q}|^2 \hspace{-0.04cm}+\hspace{-0.01cm} n_q\big)\pull - n_q \beta^{(q)}_m\pull + |\kappa^{(q)}_m|^2 \hspace{0.02cm}\big]\;.
\label{snmboson}
\end{equation}

Once the steady state is found by solving Eq.~\eqref{snmfermion} or \eqref{snmboson}, the two-site correlations can be obtained by using the product-of-modes approximation [see Eq.~\eqref{squarticapprox}],
\begin{subequations}
\begin{align}
\langle \hat{f}_i^{\dagger} \hat{f}_j \rangle =& \sum\nolimits_{m,n} \pull c_{m,i} c_{n,j}^* \langle \hat{F}_m^{\dagger} \hat{F}_n \rangle \approx \sum\nolimits_m N_m c_{m,i} c_{m,j}^*\;,
\label{sweakspcorr}\\
\text{and} \quad \langle \hat{f}_i^{\dagger} \hat{f}_j^{\dagger} \hat{f}_i \hat{f}_j \rangle 
=& \sum\nolimits_{m,n,m^{\prime},n^{\prime}} \pull c_{m,i} c_{n,j} c_{m^{\prime},i}^* c_{n^{\prime},j}^*
\langle \hat{F}_m^{\dagger} \hat{F}_n^{\dagger} \hat{F}_{m^{\prime}} \hat{F}_{n^{\prime}} \rangle
\approx |\langle \hat{f}_i^{\dagger} \hat{f}_j \rangle|^2 - n_i n_j\;.
\label{sweaknncorr}
\end{align}
\end{subequations}
The latter can be used to find the density-density correlations for both free fermions and hard-core bosons,
\begin{equation}
\langle \hat{n}_i \hat{n}_j \rangle - n_i n_j := \langle \hat{f}_i^{\dagger} \hat{f}_i \hat{f}_j^{\dagger} \hat{f}_j \rangle - n_i n_j \approx \delta_{ij} n_i - |\langle \hat{f}_i^{\dagger} \hat{f}_j \rangle|^2 \;.
\end{equation}

\section{\label{seffstrong}Effective Zeno dynamics at strong dissipation}
In Sec.~\ref{resonance} of the main text, we modeled the dynamics at strong dissipation. Here, at long times, the pump and loss sites are pinned to occupation 1 and 0, respectively, dividing the system into weakly coupled segments. In particular, we discussed that the source (or sink) can be effectively replaced by a weak correlated pump (or loss) at its neighboring sites, which can lead to striking resonant features in steady state. Here we derive this effective Zeno dynamics using the formalism developed in Ref. \cite{sPopkov2018}.%and the remaining sites undergo an effective Zeno dynamics.

We first summarize the relevant findings in Ref. \cite{sPopkov2018}. Consider a system described by a Hamiltonian $\hat{H}$ and subject to strong dissipation characterized by a rate $\gamma$, that acts only on a subspace $\mathcal{H}_0$ of the full Hilbert space $\mathcal{H} = \mathcal{H}_0 \otimes \mathcal{H}_1$. The dissipator $\mathcal{L}_0$ targets a unique (mixed) state $\smash{\hat{\psi}_0}$ in this subspace, i.e., $\smash{\mathcal{L}_0 \hat{\psi}_0 = 0}$. Then, at all times $t \gg 1/\gamma$, the density matrix $\hat{\rho}$ is well approximated as $\smash{\hat{\rho} \approx \hat{\psi}_0 \otimes \hat{\rho}_{\text{eff}}}$, where $\hat{\rho}_{\text{eff}}$ encodes the state in $\mathcal{H}_1$. The time evolution of $\hat{\rho}_{\text{eff}}$ depends on the spectrum of $\mathcal{L}_0$, which is composed of eigenvalues $\xi_{\nu}$, and left and right eigenvectors $\smash{\hat{\phi}_{\nu}}$ and $\smash{\hat{\psi}_{\nu}}$ such that $\smash{\text{Tr}_{\mathcal{H}_0} (\hat{\phi}_{\mu} \hat{\psi}_{\nu}) = \delta_{\mu,\nu}}$ ($\mu,\nu \geq 0$). In particular, one can show $d\hat{\rho}_{\text{eff}}/dt \approx -({\rm i}/\hbar) [\hat{H}_{\text{eff}}, \hat{\rho}_{\text{eff}}] + \mathcal{D}_{\text{eff}} \hat{\rho}_{\text{eff}}$ with
\begin{align}
\hat{H}_{\text{eff}} =&\; \hat{g}_0 + \sum\nolimits_{\mu,\nu>0} (\text{Im} \; Y_{\mu,\nu}) \push \hat{g}_{\mu}^{\dagger} \hat{g}_{\nu}\;,
\label{sHeffgen}\\
\text{and} \quad \mathcal{D}_{\text{eff}} \push \hat{\rho}_{\text{eff}} =&\; \sum\nolimits_{\mu,\nu>0} (\text{Re} \; Y_{\mu,\nu}) \push \big(2 \hat{g}_{\nu} \hat{\rho}_{\text{eff}} \push \hat{g}_{\mu}^{\dagger} - \{ \hat{g}_{\mu}^{\dagger} \hat{g}_{\nu}, \hat{\rho}_{\text{eff}} \} \big)\;,
\label{sDeffgen}
\end{align}
where the operators $\smash{\hat{g}_{\nu}}$ and coefficients $Y_{\mu,\nu}$ are given by
\begin{align}
\hat{g}_{\nu} &:= \text{Tr}_{\mathcal{H}_0} [(\hat{\psi}_{\nu} \otimes \hat{\mathds{1}}_{\mathcal{H}_1}) \hat{H}] \;,
\label{sdefineg}\\
\text{and} \quad Y_{\mu,\nu} &:= -\text{Tr}_{\mathcal{H}_0} (\hat{\phi}_{\mu}^{\dagger} \hat{\phi}_{\nu} \hat{\psi}_0) / \xi_{\mu}^*\;.
\label{sdefineY}
\end{align}
Note the eigenvalues $\xi_{\nu}$ scale as $\gamma$, so the effective dissipation $\mathcal{D}_{\text{eff}}$ falls off as $1/\gamma$. Also, $\smash{\hat{g}_0}$ in Eq.~\eqref{sHeffgen} is simply the Hamiltonian $\smash{\hat{H}}$ projected onto the target state $\smash{\hat{\psi}_0}$.%effective dissipation scales as $1/\gamma$, and $\smash{\hat{g}_0}$ in Eq.~\eqref{sHeffgen} is the Hamiltonian $\smash{\hat{H}}$ projected onto the target $\smash{\hat{\psi}_0}$.

\begin{table}
\caption{\label{seigtable}Eigenvalues $\smash{\xi^{p,q}_{\nu}}$, right eigenvectors $\smash{\hat{\psi}^{p,q}_{\nu}}$, and left eigenvectors $\smash{\hat{\phi}^{p,q}_{\nu}}$ of the pump and loss dissipators $\smash{\mathcal{D}[\hat{b}_p^{\dagger}]}$ and $\smash{\mathcal{D}[\hat{b}_q]}$.}
\begin{ruledtabular}
\begin{tabular}{crrrrrr}
& $\xi^p_{\nu}$ & $\smash{\hat{\psi}^p_{\nu}}$ & $\smash{\hat{\phi}^p_{\nu}}$ & $\xi^q_{\nu}$ & $\smash{\hat{\psi}^q_{\nu}}$ & $\smash{\hat{\phi}^q_{\nu}}$ \vspace{0.05cm}\\ \hline
$\nu=0$ & 0 & $\hat{n}_p$ & $\hat{\mathds{1}}_p$ & 0 & $\hat{\mathds{1}}_q - \hat{n}_q$ & $\hat{\mathds{1}}_q$\\
$\nu=1$ & $-1/2$ & $\hat{b}_p^{\dagger}$ & $\hat{b}_p$ & $-1/2$ & $\hat{b}_q$ & $\hat{b}_q^{\dagger}$\\
$\nu=2$ & $-1/2$ & $\hat{b}_p$ & $\hat{b}_p^{\dagger}$ & $-1/2$ & $\hat{b}_q^{\dagger}$ & $\hat{b}_q$ \\
$\nu=3$ & $-1$ & $\hat{\mathds{1}}_p - 2 \hat{n}_p$ & $\hat{\mathds{1}}_p - \hat{n}_p$ & $-1$ & $2\hat{n}_q - \hat{\mathds{1}}_q$ & $\hat{n}_q$
\end{tabular}
\end{ruledtabular}
\end{table}

In our qubit array, $\mathcal{L}_0 = \gamma_+ \mathcal{D}[\hat{b}_p^{\dagger}] + \gamma_- \mathcal{D}[\hat{b}_q]$, where $\mathcal{D}[\hat{b}_p^{\dagger}]$ and $\mathcal{D}[\hat{b}_q]$ are the pump and loss dissipators, respectively. (Recall, $\smash{\mathcal{D}[\hat{x}] \hat{\rho} := \hat{x} \hat{\rho} \hat{x}^{\dagger} - \{\hat{x}^{\dagger} \hat{x}, \hat{\rho} \}/2}$.) As these two act on disjoint subspaces ($q>p$), $\mathcal{L}_0$ can be diagonalized in terms of their individual eigenvalues and eigenvectors listed in Table \ref{seigtable}. The eigenvalues of $\mathcal{L}_0$ are given by $\xi_{\nu,\nu^{\prime}} \pull= \gamma_+\xi^p_{\nu} + \gamma_-\xi^q_{\nu^{\prime}}$, and the corresponding eigenvectors are $\smash{\hat{\psi}_{\nu,\nu^{\prime}} = \hat{\psi}^p_{\nu} \otimes \hat{\psi}^q_{\nu^{\prime}}}$ and $\smash{\hat{\phi}_{\nu,\nu^{\prime}} = \hat{\phi}^p_{\nu} \otimes \hat{\phi}^q_{\nu^{\prime}}}$. One can readily verify the eigenvectors are normalized such that $\smash{\text{Tr}_{\mathcal{H}_p} (\hat{\phi}_{\mu}^p \hat{\psi}_{\nu}^p) = \delta_{\mu,\nu}}$ and $\smash{\text{Tr}_{\mathcal{H}_q} (\hat{\phi}_{\mu}^q \hat{\psi}_{\nu}^q) = \delta_{\mu,\nu}}$. As expected, the target state $\smash{\hat{\psi}_{0,0}}$ describes a filled pump site and an empty loss site in the subspace $\mathcal{H}_0 = \mathcal{H}_p \otimes \mathcal{H}_q$. Using Eq.~\eqref{sdefineg}, we find
\begin{equation}
\hat{g}_{0,0} =  -J \sum\nolimits_{i=1}^{L-1} \text{Tr}_{\mathcal{H}_0} \big[ \hat{b}_p^{\dagger} \hat{b}_p \hat{b}_q \hat{b}_q^{\dagger} \big(\hat{b}_{i+1}^{\dagger} \hat{b}_i + \hat{b}_i^{\dagger} \hat{b}_{i+1} \big)\big]
= -J \Big(\sum\nolimits_{i=1}^{p-2} + \sum\nolimits_{i=p+1}^{q-2} + \sum\nolimits_{i=q+1}^{L-1}\Big) \big( \hat{b}_{i+1}^{\dagger} \hat{b}_i + \text{H.c.} \big)\;.
\label{sg00}
\end{equation}
Thus, the Hamiltonian projected onto the Zeno subspace simply describes hopping in three uncoupled segments. The only nonzero coefficients $Y_{(\mu,\mu^{\prime}),(\nu,\nu^{\prime})}$ in Eq.~\eqref{sdefineY}, with $\xi_{\mu,\mu^{\prime}}, \xi_{\nu,\nu^{\prime}} \neq 0$, are
\begin{equation}
Y_{(1,0),(1,0)} = 2/\gamma_+\;,\quad Y_{(0,1),(0,1)} = 2/\gamma_-\;, \quad \text{and} \quad Y_{(1,1),(1,1)} = 2/(\gamma_+ \pull + \gamma_-)\;.
\label{snonzeroYs}
\end{equation}
The corresponding dissipators $\smash{\hat{g}_{\nu,\nu^{\prime}}}$ in Eq.~\eqref{sdefineg} are obtained as (for $q>p$)
\begin{subequations}
\begin{align}
\hat{g}_{1,0} &= -J \sum\nolimits_{i=1}^{L-1} \text{Tr}_{\mathcal{H}_0} \big[ \hat{b}_p^{\dagger} \hat{b}_q \hat{b}_q^{\dagger} \big(\hat{b}_{i+1}^{\dagger} \hat{b}_i + \hat{b}_i^{\dagger} \hat{b}_{i+1} \big)\big] 
= -J \big[(1-\delta_{p,1})\push \hat{b}_{p-1}^{\dagger} + (1-\delta_{q,p+1})\push \hat{b}_{p+1}^{\dagger} \big]\;,
\label{sg10}\\
\hat{g}_{0,1} &= -J \sum\nolimits_{i=1}^{L-1} \text{Tr}_{\mathcal{H}_0} \big[ \hat{b}_p^{\dagger} \hat{b}_p \hat{b}_q \big(\hat{b}_{i+1}^{\dagger} \hat{b}_i + \hat{b}_i^{\dagger} \hat{b}_{i+1} \big)\big] 
= -J \big[(1-\delta_{q,p+1})\push \hat{b}_{q-1} + (1-\delta_{q,L})\push \hat{b}_{q+1} \big]\;,
\label{sg01}\\
\hat{g}_{1,1} &= -J \sum\nolimits_{i=1}^{L-1} \text{Tr}_{\mathcal{H}_0} \big[ \hat{b}_p^{\dagger} \hat{b}_q \big(\hat{b}_{i+1}^{\dagger} \hat{b}_i + \hat{b}_i^{\dagger} \hat{b}_{i+1} \big)\big] 
= -J \delta_{q,p+1} \hat{\mathds{1}}_{\mathcal{H}_1}\;.
\label{sg11}
\end{align}
\end{subequations}
Substituting these results into Eqs.~\eqref{sHeffgen} and \eqref{sDeffgen}, one finds $\smash{\hat{H}_{\text{eff}} = \hat{g}_{0,0}}$ and $\mathcal{D}_{\text{eff}} = \mathcal{D}[\hat{L}^{\text{eff}}_+] + \mathcal{D}[\hat{L}^{\text{eff}}_-]$, where
\begin{subequations}
\label{sLeff}
\begin{align}
\hat{L}^{\text{eff}}_+ &= \sqrt{\Gamma}_+ \big[(1-\delta_{p,1})\push \hat{b}_{p-1}^{\dagger} + (1-\delta_{q,p+1})\push \hat{b}_{p+1}^{\dagger} \big] \;,
\label{sLeffplus}\\
\hat{L}^{\text{eff}}_- &= \sqrt{\Gamma}_- \big[(1-\delta_{q,p+1})\push \hat{b}_{q-1} + (1-\delta_{q,L})\push \hat{b}_{q+1} \big]\;,
\label{sLeffminus}
\end{align}
\end{subequations}
with rates $\Gamma_{\pm} := 4J^2/\gamma_{\pm}$. Therefore, the source and sink generates correlation between its neighboring sites through a second-order process, dissipatively coupling the segments in Eq.~\eqref{sg00}. For pump and loss of free fermions instead of hard-core bosons, one finds the same expressions with $\smash{\hat{b}}$'s replaced by $\smash{\hat{f}}$'s in Eqs.~\eqref{sLeff}.

\section{\label{srateeqsstrong}Rate equations for strong dissipation}
We showed above that the qubit array reduces to weakly coupled segments at strong dissipation. Since this coupling is small compared to tunneling, any off-resonant modes become uncorrelated at long times, to a good approximation, as discussed in Sec.~\ref{resonance} of the main article. This is similar to what happens at weak dissipation (see Sec.~\ref{srateeqsweak}). However, the modes that are resonant in neighboring segments can remain coherent, producing surprising steady states. Here we derive the approximate rate equations that explain these features.

The pump and loss divide the system into three segments as in Eq.~\eqref{sg00}. The single-particle modes in segment $\nu$, with $L_{\nu}$ sites, have the form $\smash{\hat{F}^{(\nu)}_{m_{\nu}} \pull=\pull \sum_{j_{\nu}=1}^{L_{\nu}}\pull c^{(\nu)}_{m_{\nu},j_{\nu}} \hat{f}^{(\nu)}_{j_{\nu}}}$, $m_{\nu} = 1,\dots,L_{\nu}$. Here %$\smash{\hat{f}^{(1)}_{j_1} \pull:= \hat{f}_{j_1}}$, $\smash{\hat{f}^{(2)}_{j_2} \pull:= \hat{f}_{p+ j_2}}$, and $\smash{\hat{f}^{(3)}_{j_3} \pull:= \hat{f}_{q+j_3}}$.
\begin{equation}
\hat{f}^{(1)}_{j_1} \pull:= \hat{f}_{j_1}\push, \quad  \hat{f}^{(2)}_{j_2} \pull:= \hat{f}_{p+ j_2} \;, \quad \text{and}\;\;\; \hat{f}^{(3)}_{j_3} \pull:= \hat{f}_{q+j_3}\;.
\end{equation}
Note, in this section we use $\nu$ to label the segments, and not eigenvalues as in the last section! The modes in neighboring segments are coupled through the dissipators in Eqs.~\eqref{sLeff}. For simplicity, we assume all three segments are present, i.e., $1<p<q \pull-\pull 1<L \pull-\pull 1$, which gives
\begin{equation}
\hat{L}^{\text{eff}}_+ = \sqrt{\Gamma_+} (\hat{b}_{p-1}^{\dagger} \pull+ \hat{b}_{p+1}^{\dagger})
\quad \text{and} \quad \hat{L}^{\text{eff}}_- = \sqrt{\Gamma_-} (\hat{b}_{q-1} \pull+ \hat{b}_{q+1})\;.
\label{sLeffsimple}
\end{equation}
To find the equations of motion, it is useful to write the dissipators in terms of the modes. To this end, we first invert the unitary coefficients $\smash{c^{(\nu)}_{m_{\nu},j_{\nu}}}$ to find $\smash{\hat{f}^{(\nu)}_{j_{\nu}}}$, then attach appropriate string operators according to Eq.~\eqref{sJordanWigner}, yielding
\begin{subequations}
\label{seffectivedissipators}
\begin{align}
\hat{b}_{p-1} &= (-1)^{\hat{\mathcal{N}}_1} \sum\nolimits_{m_1} \pull v^{(1)*}_{m_1} \hat{F}^{(1)}_{m_1}\;, \hspace{0.82cm} 
\hat{b}_{q-1} = -(-1)^{\hat{\mathcal{N}}_{1,2}} \sum\nolimits_{m_2} \pull v^{(2)*}_{m_2} \hat{F}^{(2)}_{m_2}\;, 
\\
\hat{b}_{p+1} &= -(-1)^{\hat{\mathcal{N}}_1} \sum\nolimits_{m_2} \pull u^{(2)*}_{m_2} \hat{F}^{(2)}_{m_2}\;, \hspace{0.56cm} 
\hat{b}_{q+1} = -(-1)^{\hat{\mathcal{N}}_{1,2}} \sum\nolimits_{m_3} \pull u^{(3)*}_{m_3} \hat{F}^{(3)}_{m_3}\;,
\end{align}
\end{subequations}
%\begin{subequations}
%\begin{align}
%\hat{b}_{p-1} &= (-1)^{\hat{\mathcal{N}}_1} \sum\nolimits_{m_1} \pull v^{(1)*}_{m_1} \hat{F}^{(1)}_{m_1}\;,
%\label{sbpminus}\\
%\hat{b}_{p+1} &= -(-1)^{\hat{\mathcal{N}}_1} \sum\nolimits_{m_2} \pull u^{(2)*}_{m_2} \hat{F}^{(2)}_{m_2}\;,
%\label{sbpplus}\\
%\hat{b}_{q-1} &= -(-1)^{\hat{\mathcal{N}}_{1,2}} \sum\nolimits_{m_2} \pull v^{(2)*}_{m_2} \hat{F}^{(2)}_{m_2}\;,
%\label{sbqminus}\\
%\text{and} \quad \hat{b}_{q+1} &= -(-1)^{\hat{\mathcal{N}}_{1,2}} \sum\nolimits_{m_3} \pull u^{(3)*}_{m_3} \hat{F}^{(3)}_{m_3}\;,
%\end{align}
%\end{subequations}
where $\smash{u^{(\nu)}_{m_{\nu}}:=c^{(\nu)}_{m_{\nu},1}}$, $\smash{v^{(\nu)}_{m_{\nu}}:=c^{(\nu)}_{m_{\nu},L_{\nu}}}$, $\mathcal{N}_1$ is the total occupation in the first segment, and $\mathcal{N}_{1,2}$ is the total occupation in the first two segments. The extra minus signs arise from the filled pump site and do not affect the physics. Note the convention for $\mathcal{N}_1$ is different from that used in Sec.~\ref{srateeqsweak}. As explained in the main text, the steady state is set by the occupations $\smash{N^{(\nu)}_{m_{\nu}} := \langle \hat{N}^{(\nu)}_{m_{\nu}} \rangle = \langle \hat{F}^{(\nu) \dagger}_{m_{\nu}} \hat{F}^{(\nu)}_{m_{\nu}}\rangle}$ and the correlations $\smash{\langle \hat{F}^{(\nu) \dagger}_{m_{\nu}} \hat{F}^{(\nu+1)}_{m_{\nu+1}}\rangle}$. We find the rate equations for $\smash{N^{(3)}_{m_{3}}}$ and $\smash{\langle \hat{F}^{(2) \dagger}_{m_{2}} \hat{F}^{(3)}_{m_{3}}\rangle}$ from first principles, and later work out those for the other segments by symmetry.

The equation of motion for the occupations can be found from Eq.~\eqref{sobseom},
\begin{equation}
\dot{N}^{(3)}_{m_3} = -\text{Re}\push \langle \hat{L}^{\text{eff} \dagger}_+ [\hat{L}^{\text{eff}}_+, \hat{N}^{(3)}_{m_3}] \rangle -\text{Re}\push \langle \hat{L}^{\text{eff} \dagger}_- [\hat{L}^{\text{eff}}_-, \hat{N}^{(3)}_{m_3}] \rangle\;.
\label{sn3def}
\end{equation}
Using Eqs.~\eqref{seffectivedissipators}, one finds $\smash{[\hat{L}^{\text{eff}}_+, \hat{N}^{(3)}_{m_3}] = [\hat{b}_{q-1},\hat{N}^{(3)}_{m_3}] = 0}$ and $\smash{[\hat{b}_{q+1},\hat{N}^{(3)}_{m_3}] = -(-1)^{\hat{\mathcal{N}}_{1,2}} u^{(3)*}_{m_3} \hat{F}^{(3)}_{m_3}}$, which give
\begin{equation}
\dot{N}^{(3)}_{m_3} 
= -\Gamma_- \push \text{Re}\; u^{(3)*}_{m_3} \Big[ 
\sum\nolimits_{m_2} \pull v^{(2)}_{m_2} \push \langle \hat{F}^{(2) \dagger}_{m_{2}} \hat{F}^{(3)}_{m_{3}}\rangle 
+ \sum\nolimits_{m_3^{\prime}} \pull u^{(3)}_{m_3^{\prime}} \push \langle \hat{F}^{(3) \dagger}_{m_{3}^{\prime}} \hat{F}^{(3)}_{m_{3}}\rangle \Big] \;.
\label{sn3expand}
\end{equation}
Next, we make the approximation that all off-resonant modes are uncorrelated and that the spectrum is nondegenerate in a given segment, i.e., $\smash{\langle \hat{F}^{(\nu)\dagger}_{m_{\nu}}\pull \hat{F}^{(\nu)}_{m_{\nu}^{\prime}}\rangle \pull\approx \delta_{m_{\nu},m_{\nu}^{\prime}} N^{(\nu)}_{m_{\nu}}}$ as in Sec.~\ref{srateeqsweak}, and $\smash{\langle \hat{F}^{(\nu)\dagger}_{m_{\nu}}\pull \hat{F}^{(\nu+1)}_{m_{\nu+1}}\rangle \pull\approx\pull \Lambda^{(\nu,\nu+1)}_{m_{\nu},m_{\nu+1}} T^{(\nu,\nu+1)}_{m_{\nu},m_{\nu+1}}}$, where $\smash{\Lambda^{(\nu,\nu^{\prime})}_{m_{\nu},m_{\nu^{\prime}}}}$ is 1 for resonant modes and 0 otherwise. Substituting these into Eq.~\eqref{sn3expand} yields
\begin{equation}
\dot{N}^{(3)}_{m_3} \approx -\Gamma_{-} \Big[|u^{(3)}_{m_3}|^2 N^{(3)}_{m_3} +\text{Re}\push \sum\nolimits_{m_2}\pull \Lambda^{(2,3)}_{m_2,m_3} v^{(2)}_{m_2} u^{(3)*}_{m_3} T^{(2,3)}_{m_2,m_3}\Big]\;.
\label{sn3final}
\end{equation}
The same rate equation is obtained if the hard-core bosons were replaced by free fermions in Eq.~\eqref{sLeffsimple}.

To find the rate equation for the correlations, we first generalize Eq.~\eqref{sobseom} for a non-Hermitian operator $\smash{\hat{X}}$. Using $\smash{ \langle\hat{X}\rangle \pull=\pull \text{Tr}(\hat{X}\hat{\rho})}$ in Eq.~\eqref{smastereqn}, along with the effective Hamiltonians and dissipators, one finds
\begin{equation}
\frac{d\langle\hat{X}\rangle}{dt} 
= \frac{{\rm i}}{\hbar} \langle [\hat{H}_{\text{eff}},\hat{X}] \rangle 
+ \frac{1}{2}\push \sum\nolimits_{\alpha=\pm} \langle [\hat{L}_{\alpha}^{\text{eff}\dagger}, \hat{X}] \hat{L}_{\alpha}^{\text{eff}} \rangle - \langle \hat{L}_{\alpha}^{\text{eff}\dagger} [\hat{L}_{\alpha}^{\text{eff}}, \hat{X}]\rangle\;.
\label{sobseomgeneral}
\end{equation}
%\begin{equation}
%d\langle \hat{X} \rangle / dt 
%= ({\rm i}/\hbar) \langle [\hat{H},\hat{X}] \rangle 
%+ (1/2) \sum\nolimits_{\alpha=\pm} \langle [\hat{L}_{\alpha}^{\dagger}, \hat{X}] \hat{L}_{\alpha} \rangle - \langle \hat{L}_{\alpha}^{\dagger} [\hat{L}_{\alpha}, \hat{X}]\rangle\;.
%\label{sobseomgeneral}
%\end{equation}
We consider the correlation between two resonant modes $m_2$ and $m_3$, i.e., $\smash{\hat{X} = \hat{F}^{(2) \dagger}_{m_{2}} \hat{F}^{(3)}_{m_{3}}}$ with $\Lambda^{(2,3)}_{m_2,m_3} = 1$. Since the two modes have equal energies, $\smash{[\hat{H}_{\text{eff}},\hat{X}]=0}$. Further, using Eqs.~\eqref{sLeffsimple} and \eqref{seffectivedissipators} gives $\smash{[\hat{L}^{\text{eff}}_+,\hat{X}]=[\hat{b}_{p-1},\hat{X}]=0}$ and $\smash{[\hat{b}_{p+1},\hat{X}] = -(-1)^{\hat{\mathcal{N}}_1} u^{(2)*}_{m_2} \hat{F}^{(3)}_{m_3}}$. Noting that $\smash{\big\{(-1)^{\hat{\mathcal{N}}_1}, \hat{F}^{(1)}_{m_1}\big\}=0}$, and $\big[(-1)^{\hat{\mathcal{N}}_1}, \hat{F}^{(\nu)}_{m_{\nu}}\big]=0$ for $\nu=2,3$, we obtain
\begin{align}
\nonumber \langle [\hat{L}^{\text{eff}\dagger}_+ , \hat{X}] \hat{L}^{\text{eff}}_+ \rangle
- \langle \hat{L}_{+}^{\text{eff}\dagger} [\hat{L}_{+}^{\text{eff}}, \hat{X}]\rangle
&= -\Gamma_+ u^{(2)*}_{m_2} \Big[ \sum\nolimits_{m_1} \pull\pull v^{(1)}_{m_1} \langle \hat{F}^{(1)\dagger}_{m_1} \hat{F}^{(3)}_{m_3} \rangle 
+ \sum\nolimits_{m_2^{\prime}} \pull\pull u^{(2)}_{m_2^{\prime}} \langle \hat{F}^{(2)\dagger}_{m_2^{\prime}} \hat{F}^{(3)}_{m_3} \rangle \Big] \\
& \approx -\Gamma_+ |u^{(2)}_{m_2}|^2 \push T^{(2,3)}_{m_2,m_3}\;.
\label{sT23plus}
\end{align}
Similarly, for the loss dissipator, we use $\smash{\big[(-1)^{\hat{\mathcal{N}}_{1,2}}, \hat{F}^{(3)}_{m_{3}}\big]=0}$, and $\smash{\big\{(-1)^{\hat{\mathcal{N}}_{1,2}}, \hat{F}^{(\nu)}_{m_{\nu}}\big\}=0}$ for $\nu=1,2$, which gives
\begin{subequations}
\begin{align}
[\hat{L}^{\text{eff}}_-,\hat{X}] 
&= -\sqrt{\Gamma_-} (-1)^{\hat{\mathcal{N}}_{1,2}} \big[v^{(2)*}_{m_2} \hat{F}^{(3)}_{m_3} + 2 \hat{X} \big(\hat{f}_{q-1} \pull+ \hat{f}_{q+1}\big)\big] \;,
\label{sLminusX}\\
[\hat{L}^{\text{eff}\dagger}_-,\hat{X}] 
&= \sqrt{\Gamma_-} \big[ u^{(3)}_{m_3} \hat{F}^{(2)\dagger}_{m_2} + 2 \big(\hat{f}_{q-1}^{\dagger} \pull+ \hat{f}_{q+1}^{\dagger}\big) \hat{X} \big] (-1)^{\hat{\mathcal{N}}_{1,2}}\;.
\label{sLminusdagX}
\end{align}
\end{subequations}
Combining these results with Eqs.~\eqref{seffectivedissipators} yields
\begin{equation}
\langle [\hat{L}^{\text{eff}\dagger}_- , \hat{X}] \hat{L}^{\text{eff}}_- \rangle
- \langle \hat{L}_{-}^{\text{eff}\dagger} [\hat{L}_{-}^{\text{eff}}, \hat{X}]\rangle
\approx -\Gamma_- \big[ \big( |v^{(2)}_{m_2}|^2 + |u^{(3)}_{m_3}|^2 \big) T^{(2,3)}_{m_2,m_3}
+ v^{(2)*}_{m_2} u^{(3)}_{m_3} \big( N^{(2)}_{m_2} + N^{(3)}_{m_3} \big)\big]
+ 4 \langle \hat{L}^{\text{eff}\dagger}_- \hat{X} \hat{L}^{\text{eff}}_- \rangle \;.
\label{sT23minus}
\end{equation}
Substituting Eqs.~\eqref{sT23plus} and \eqref{sT23minus} into Eq.~\eqref{sobseomgeneral}, we find the rate equation
\begin{equation}
\dot{T}^{(2,3)}_{m_2,m_3} \approx -\Gamma_{\pull f} \push T^{(2,3)}_{m_2,m_3} 
+ 2 \push \zeta_b \push \langle \hat{L}^{\text{eff}\dagger}_- \hat{F}^{(2)\dagger}_{m_2} \hat{F}^{(3)}_{m_3} \hat{L}^{\text{eff}}_- \rangle
- \Gamma_{\pull -} v^{(2)*}_{m_2} u^{(3)}_{m_3} \big[N^{(2)}_{m_2} +\pull N^{(3)}_{m_3} \big]/2 \;,
\label{sT23final}
\end{equation}
where $\Gamma_{\pull f} := \frac{\Gamma_{+}}{2} |u^{(2)}_{m_2}|^2 + \frac{\Gamma_{-}}{2} \big[|v^{(2)}_{m_2}|^2 + |u^{(3)}_{m_3}|^2\big]$, and $\zeta_b=1$ for the hard-core bosons we have been considering. For free fermion dissipators, one obtains the same equation without the quartic term, i.e., $\zeta_b=0$.

The quartic term can be approximated by pairwise contractions similar to Eq.~\eqref{squarticapprox}, yielding
\begin{equation}
\langle \hat{L}^{\text{eff}\dagger}_- \hat{F}^{(2)\dagger}_{m_2} \hat{F}^{(3)}_{m_3} \hat{L}^{\text{eff}}_- \rangle
\approx - \langle \hat{L}^{\text{eff}\dagger}_- \hat{L}^{\text{eff}}_- \rangle \push T^{(2,3)}_{m_2,m_3}
+ \Gamma_- \big\langle \hat{F}^{(2)\dagger}_{m_2} \big(\hat{f}_{q-1} \pull + \hat{f}_{q+1}\big) \big\rangle
\big\langle \big(\hat{f}_{q-1}^{\dagger} \pull + \hat{f}_{q+1}^{\dagger}\big) \hat{F}^{(3)}_{m_3} \big\rangle \;.
\label{squartics}
\end{equation}
Using this result in Eq.~\eqref{sT23final}, we find the decay rate of correlations is effectively enhanced for the hard-core bosons,
\begin{equation}
\Gamma_b \approx \Gamma_f + 2 \langle \hat{L}^{\text{eff}\dagger}_- \hat{L}^{\text{eff}}_- \rangle \;,
\label{sGammab}
\end{equation}
which dampens the resonant features in steady state. When the correlations are small compared to 1, the dominant correction in Eq.~\eqref{squartics} comes from this decay rate, $\smash{\langle \hat{L}^{\text{eff}\dagger}_- \hat{L}^{\text{eff}}_- \rangle \approx \Gamma_- n_{q-1} \sim O(\Gamma_-)}$. On the other hand, $\Gamma_{\pull f} \sim O(\Gamma_{\pm}/l)$, where $l=\text{min}(L_2,L_3)$. Thus, for large systems, $\Gamma_{b} / \Gamma_{\pull f} \sim O(l)$.

The rate equations for the other segments can be found by symmetry from Eqs.~\eqref{sn3final} and \eqref{sT23final}. In particular, we exchange segments $1 \leftrightarrow 3$, rates $\Gamma_+ \leftrightarrow \Gamma_-$, amplitudes $u \leftrightarrow v^*$, and particles with holes ($\smash{\hat{F} \leftrightarrow \hat{F}^{\dagger}}$) to obtain
\begin{align}
\dot{N}^{(1)}_{m_1} &\approx \Gamma_{+} \Big[|v^{(1)}_{m_1}|^2 \bar{N}^{(1)}_{m_1} -\text{Re}\push \sum\nolimits_{m_2}\pull \Lambda^{(1,2)}_{m_1,m_2} v^{(1)}_{m_1} u^{(2)*}_{m_2} T^{(1,2)}_{m_1,m_2}\Big]\;,
\label{sNm1}\\
\text{and} \quad \dot{T}^{(1,2)}_{m_1,m_2} &\approx -\Gamma_{\pull f}^{\prime} \push T^{(1,2)}_{m_1,m_2} 
+ 2 \push \zeta_b \push \langle \hat{L}^{\text{eff}\dagger}_+ \hat{F}^{(1)\dagger}_{m_1} \hat{F}^{(2)}_{m_2} \hat{L}^{\text{eff}}_+ \rangle
+ \Gamma_{\pull +} v^{(1)*}_{m_1} u^{(2)}_{m_2} \big[\bar{N}^{(1)}_{m_1} +\pull \bar{N}^{(2)}_{m_2} \big]/2\;,
\label{sT12}
\end{align}
where $\Gamma_{\pull f}^{\prime} := \frac{\Gamma_{+}}{2} \big[|v^{(1)}_{m_1}|^2 + |u^{(2)}_{m_2}|^2\big] + \frac{\Gamma_{-}}{2} |v^{(2)}_{m_2}|^2 $, and recall that $\bar{x}:=1-x$. The quartic term can again be approximated by pairwise contractions, yielding an increased decay rate for the bosons, $\smash{\Gamma_b^{\prime} \approx \Gamma_{\pull f}^{\prime} + 2 \langle \hat{L}^{\text{eff}\dagger}_+ \hat{L}^{\text{eff}}_+ \rangle \approx \Gamma_+ \bar{n}_{p+1}}$. The rate equation for $\smash{\hat{N}^{(2)}_{m_2}}$ is found by combining those for $\smash{\hat{N}^{(1)}_{m_1}}$ and $\smash{\hat{N}^{(3)}_{m_3}}$, and making suitable exchanges, which give
\begin{align}
\nonumber \dot{N}^{(2)}_{m_2} \approx&\; \Gamma_{+} \Big[|u^{(2)}_{m_2}|^2 \bar{N}^{(2)}_{m_2} -\text{Re}\push \sum\nolimits_{m_1}\pull \Lambda^{(1,2)}_{m_1,m_2} v^{(1)}_{m_1} u^{(2)*}_{m_2} T^{(1,2)}_{m_1,m_2}\Big]
\\
&\; - \Gamma_{-} \Big[|v^{(2)}_{m_2}|^2 N^{(2)}_{m_2} +\text{Re}\push \sum\nolimits_{m_3}\pull \Lambda^{(2,3)}_{m_2,m_3} v^{(2)}_{m_2} u^{(3)*}_{m_3} T^{(2,3)}_{m_2,m_3}\Big]\;.
\label{sn2}
\end{align}
Equations \eqref{sn3final}, \eqref{sNm1}, and \eqref{sn2} show that, without any resonance ($\Lambda=0$), $\smash{N^{(1)}_{m_1} \to 1}$ and $\smash{N^{(3)}_{m_3} \to 0}$ in steady state, while $\smash{\hat{N}^{(2)}_{m_2}}$ approaches a fraction. When resonances are present, these can be dramatically altered by the coupling to the coherences in Eqs.~\eqref{sT23final} and \eqref{sT12}, producing surprising density-wave order as shown in the main text.

%\bibliography{references_supplemental}

\begin{thebibliography}{48}%
\makeatletter
\providecommand \@ifxundefined [1]{%
 \@ifx{#1\undefined}
}%
\providecommand \@ifnum [1]{%
 \ifnum #1\expandafter \@firstoftwo
 \else \expandafter \@secondoftwo
 \fi
}%
\providecommand \@ifx [1]{%
 \ifx #1\expandafter \@firstoftwo
 \else \expandafter \@secondoftwo
 \fi
}%
\providecommand \natexlab [1]{#1}%
\providecommand \enquote  [1]{``#1''}%
\providecommand \bibnamefont  [1]{#1}%
\providecommand \bibfnamefont [1]{#1}%
\providecommand \citenamefont [1]{#1}%
\providecommand \href@noop [0]{\@secondoftwo}%
\providecommand \href [0]{\begingroup \@sanitize@url \@href}%
\providecommand \@href[1]{\@@startlink{#1}\@@href}%
\providecommand \@@href[1]{\endgroup#1\@@endlink}%
\providecommand \@sanitize@url [0]{\catcode `\\12\catcode `\$12\catcode
  `\&12\catcode `\#12\catcode `\^12\catcode `\_12\catcode `\%12\relax}%
\providecommand \@@startlink[1]{}%
\providecommand \@@endlink[0]{}%
\providecommand \url  [0]{\begingroup\@sanitize@url \@url }%
\providecommand \@url [1]{\endgroup\@href {#1}{\urlprefix }}%
\providecommand \urlprefix  [0]{URL }%
\providecommand \Eprint [0]{\href }%
\providecommand \doibase [0]{http://dx.doi.org/}%
\providecommand \selectlanguage [0]{\@gobble}%
\providecommand \bibinfo  [0]{\@secondoftwo}%
\providecommand \bibfield  [0]{\@secondoftwo}%
\providecommand \translation [1]{[#1]}%
\providecommand \BibitemOpen [0]{}%
\providecommand \bibitemStop [0]{}%
\providecommand \bibitemNoStop [0]{.\EOS\space}%
\providecommand \EOS [0]{\spacefactor3000\relax}%
\providecommand \BibitemShut  [1]{\csname bibitem#1\endcsname}%
\let\auto@bib@innerbib\@empty
%</preamble>
\bibitem [{\citenamefont {Schlosshauer}(2019)}]{Schlosshauer2019}%
  \BibitemOpen
  \bibfield  {author} {\bibinfo {author} {\bibfnamefont {M.}~\bibnamefont
  {Schlosshauer}},\ }\bibfield  {title} {\enquote {\bibinfo {title} {Quantum
  decoherence},}\ }\href {\doibase 10.1016/j.physrep.2019.10.001} {\bibfield
  {journal} {\bibinfo  {journal} {Phys. Rep.}\ }\textbf {\bibinfo {volume}
  {831}},\ \bibinfo {pages} {1} (\bibinfo {year} {2019})}\BibitemShut {NoStop}%
\bibitem [{\citenamefont {M{\"u}ller}\ \emph {et~al.}(2012)\citenamefont
  {M{\"u}ller}, \citenamefont {Diehl}, \citenamefont {Pupillo},\ and\
  \citenamefont {Zoller}}]{Mueller2012}%
  \BibitemOpen
  \bibfield  {author} {\bibinfo {author} {\bibfnamefont {M.}~\bibnamefont
  {M{\"u}ller}}, \bibinfo {author} {\bibfnamefont {S.}~\bibnamefont {Diehl}},
  \bibinfo {author} {\bibfnamefont {G.}~\bibnamefont {Pupillo}}, \ and\
  \bibinfo {author} {\bibfnamefont {P.}~\bibnamefont {Zoller}},\ }\bibfield
  {title} {\enquote {\bibinfo {title} {Engineered open systems and quantum
  simulations with atoms and ions},}\ }\href {\doibase
  10.1016/b978-0-12-396482-3.00001-6} {\bibfield  {journal} {\bibinfo
  {journal} {Adv. At. Mol. Opt. Phys.}\ }\textbf {\bibinfo {volume} {61}},\
  \bibinfo {pages} {1} (\bibinfo {year} {2012})}\BibitemShut {NoStop}%
\bibitem [{\citenamefont {Diehl}\ \emph {et~al.}(2011)\citenamefont {Diehl},
  \citenamefont {Rico}, \citenamefont {Baranov},\ and\ \citenamefont
  {Zoller}}]{Diehl2011}%
  \BibitemOpen
  \bibfield  {author} {\bibinfo {author} {\bibfnamefont {S.}~\bibnamefont
  {Diehl}}, \bibinfo {author} {\bibfnamefont {E.}~\bibnamefont {Rico}},
  \bibinfo {author} {\bibfnamefont {M.~A.}\ \bibnamefont {Baranov}}, \ and\
  \bibinfo {author} {\bibfnamefont {P.}~\bibnamefont {Zoller}},\ }\bibfield
  {title} {\enquote {\bibinfo {title} {Topology by dissipation in atomic
  quantum wires},}\ }\href {\doibase 10.1038/nphys2106} {\bibfield  {journal}
  {\bibinfo  {journal} {Nat. Phys.}\ }\textbf {\bibinfo {volume} {7}},\
  \bibinfo {pages} {971} (\bibinfo {year} {2011})}\BibitemShut {NoStop}%
\bibitem [{\citenamefont {Lin}\ \emph {et~al.}(2013)\citenamefont {Lin},
  \citenamefont {Gaebler}, \citenamefont {Reiter}, \citenamefont {Tan},
  \citenamefont {Bowler}, \citenamefont {S{\o}rensen}, \citenamefont
  {Leibfried},\ and\ \citenamefont {Wineland}}]{Lin2013}%
  \BibitemOpen
  \bibfield  {author} {\bibinfo {author} {\bibfnamefont {Y.}~\bibnamefont
  {Lin}}, \bibinfo {author} {\bibfnamefont {J.~P.}\ \bibnamefont {Gaebler}},
  \bibinfo {author} {\bibfnamefont {F.}~\bibnamefont {Reiter}}, \bibinfo
  {author} {\bibfnamefont {T.~R.}\ \bibnamefont {Tan}}, \bibinfo {author}
  {\bibfnamefont {R.}~\bibnamefont {Bowler}}, \bibinfo {author} {\bibfnamefont
  {A.~S.}\ \bibnamefont {S{\o}rensen}}, \bibinfo {author} {\bibfnamefont
  {D.}~\bibnamefont {Leibfried}}, \ and\ \bibinfo {author} {\bibfnamefont
  {D.~J.}\ \bibnamefont {Wineland}},\ }\bibfield  {title} {\enquote {\bibinfo
  {title} {Dissipative production of a maximally entangled steady state of two
  quantum bits},}\ }\href {\doibase 10.1038/nature12801} {\bibfield  {journal}
  {\bibinfo  {journal} {Nature (London)}\ }\textbf {\bibinfo {volume} {504}},\
  \bibinfo {pages} {415} (\bibinfo {year} {2013})}\BibitemShut {NoStop}%
\bibitem [{\citenamefont {Carr}\ and\ \citenamefont
  {Saffman}(2013)}]{Carr2013}%
  \BibitemOpen
  \bibfield  {author} {\bibinfo {author} {\bibfnamefont {A.~W.}\ \bibnamefont
  {Carr}}\ and\ \bibinfo {author} {\bibfnamefont {M.}~\bibnamefont {Saffman}},\
  }\bibfield  {title} {\enquote {\bibinfo {title} {Preparation of entangled and
  antiferromagnetic states by dissipative rydberg pumping},}\ }\href {\doibase
  10.1103/physrevlett.111.033607} {\bibfield  {journal} {\bibinfo  {journal}
  {Phys. Rev. Lett.}\ }\textbf {\bibinfo {volume} {111}},\ \bibinfo {pages}
  {033607} (\bibinfo {year} {2013})}\BibitemShut {NoStop}%
\bibitem [{\citenamefont {Sieberer}\ \emph {et~al.}(2016)\citenamefont
  {Sieberer}, \citenamefont {Buchhold},\ and\ \citenamefont
  {Diehl}}]{Sieberer2016}%
  \BibitemOpen
  \bibfield  {author} {\bibinfo {author} {\bibfnamefont {L.~M.}\ \bibnamefont
  {Sieberer}}, \bibinfo {author} {\bibfnamefont {M.}~\bibnamefont {Buchhold}},
  \ and\ \bibinfo {author} {\bibfnamefont {S.}~\bibnamefont {Diehl}},\
  }\bibfield  {title} {\enquote {\bibinfo {title} {Keldysh field theory for
  driven open quantum systems},}\ }\href {\doibase
  10.1088/0034-4885/79/9/096001} {\bibfield  {journal} {\bibinfo  {journal}
  {Rep. Prog. Phys.}\ }\textbf {\bibinfo {volume} {79}},\ \bibinfo {pages}
  {096001} (\bibinfo {year} {2016})}\BibitemShut {NoStop}%
\bibitem [{\citenamefont {Kordas}\ \emph {et~al.}(2015)\citenamefont {Kordas},
  \citenamefont {Witthaut}, \citenamefont {Buonsante}, \citenamefont {Vezzani},
  \citenamefont {Burioni}, \citenamefont {Karanikas},\ and\ \citenamefont
  {Wimberger}}]{Kordas2015}%
  \BibitemOpen
  \bibfield  {author} {\bibinfo {author} {\bibfnamefont {G.}~\bibnamefont
  {Kordas}}, \bibinfo {author} {\bibfnamefont {D.}~\bibnamefont {Witthaut}},
  \bibinfo {author} {\bibfnamefont {P.}~\bibnamefont {Buonsante}}, \bibinfo
  {author} {\bibfnamefont {A.}~\bibnamefont {Vezzani}}, \bibinfo {author}
  {\bibfnamefont {R.}~\bibnamefont {Burioni}}, \bibinfo {author} {\bibfnamefont
  {A.~I.}\ \bibnamefont {Karanikas}}, \ and\ \bibinfo {author} {\bibfnamefont
  {S.}~\bibnamefont {Wimberger}},\ }\bibfield  {title} {\enquote {\bibinfo
  {title} {The dissipative {B}ose-{H}ubbard model},}\ }\href {\doibase
  10.1140/epjst/e2015-02528-2} {\bibfield  {journal} {\bibinfo  {journal} {Eur.
  Phys. J. Special Topics}\ }\textbf {\bibinfo {volume} {224}},\ \bibinfo
  {pages} {2127} (\bibinfo {year} {2015})}\BibitemShut {NoStop}%
\bibitem [{\citenamefont {Verstraete}\ \emph {et~al.}(2009)\citenamefont
  {Verstraete}, \citenamefont {Wolf},\ and\ \citenamefont
  {Cirac}}]{Verstraete2009}%
  \BibitemOpen
  \bibfield  {author} {\bibinfo {author} {\bibfnamefont {F.}~\bibnamefont
  {Verstraete}}, \bibinfo {author} {\bibfnamefont {M.~M.}\ \bibnamefont
  {Wolf}}, \ and\ \bibinfo {author} {\bibfnamefont {J.~I.}\ \bibnamefont
  {Cirac}},\ }\bibfield  {title} {\enquote {\bibinfo {title} {Quantum
  computation and quantum-state engineering driven by dissipation},}\ }\href
  {\doibase 10.1038/nphys1342} {\bibfield  {journal} {\bibinfo  {journal} {Nat.
  Phys.}\ }\textbf {\bibinfo {volume} {5}},\ \bibinfo {pages} {633} (\bibinfo
  {year} {2009})}\BibitemShut {NoStop}%
\bibitem [{\citenamefont {Cazalilla}\ \emph {et~al.}(2011)\citenamefont
  {Cazalilla}, \citenamefont {Citro}, \citenamefont {Giamarchi}, \citenamefont
  {Orignac},\ and\ \citenamefont {Rigol}}]{Cazalilla2011}%
  \BibitemOpen
  \bibfield  {author} {\bibinfo {author} {\bibfnamefont {M.~A.}\ \bibnamefont
  {Cazalilla}}, \bibinfo {author} {\bibfnamefont {R.}~\bibnamefont {Citro}},
  \bibinfo {author} {\bibfnamefont {T.}~\bibnamefont {Giamarchi}}, \bibinfo
  {author} {\bibfnamefont {E.}~\bibnamefont {Orignac}}, \ and\ \bibinfo
  {author} {\bibfnamefont {M.}~\bibnamefont {Rigol}},\ }\bibfield  {title}
  {\enquote {\bibinfo {title} {One dimensional bosons: From condensed matter
  systems to ultracold gases},}\ }\href {\doibase 10.1103/revmodphys.83.1405}
  {\bibfield  {journal} {\bibinfo  {journal} {Rev. Mod. Phys.}\ }\textbf
  {\bibinfo {volume} {83}},\ \bibinfo {pages} {1405} (\bibinfo {year}
  {2011})}\BibitemShut {NoStop}%
\bibitem [{\citenamefont {Ma}\ \emph {et~al.}(2019)\citenamefont {Ma},
  \citenamefont {Saxberg}, \citenamefont {Owens}, \citenamefont {Leung},
  \citenamefont {Lu}, \citenamefont {Simon},\ and\ \citenamefont
  {Schuster}}]{Ma2019}%
  \BibitemOpen
  \bibfield  {author} {\bibinfo {author} {\bibfnamefont {R.}~\bibnamefont
  {Ma}}, \bibinfo {author} {\bibfnamefont {B.}~\bibnamefont {Saxberg}},
  \bibinfo {author} {\bibfnamefont {C.}~\bibnamefont {Owens}}, \bibinfo
  {author} {\bibfnamefont {N.}~\bibnamefont {Leung}}, \bibinfo {author}
  {\bibfnamefont {Y.}~\bibnamefont {Lu}}, \bibinfo {author} {\bibfnamefont
  {J.}~\bibnamefont {Simon}}, \ and\ \bibinfo {author} {\bibfnamefont {D.~I.}\
  \bibnamefont {Schuster}},\ }\bibfield  {title} {\enquote {\bibinfo {title} {A
  dissipatively stabilized {M}ott insulator of photons},}\ }\href {\doibase
  10.1038/s41586-019-0897-9} {\bibfield  {journal} {\bibinfo  {journal} {Nature
  (London)}\ }\textbf {\bibinfo {volume} {566}},\ \bibinfo {pages} {51}
  (\bibinfo {year} {2019})}\BibitemShut {NoStop}%
\bibitem [{\citenamefont {Prosen}(2008)}]{Prosen2008}%
  \BibitemOpen
  \bibfield  {author} {\bibinfo {author} {\bibfnamefont {T.}~\bibnamefont
  {Prosen}},\ }\bibfield  {title} {\enquote {\bibinfo {title} {Third
  quantization: a general method to solve master equations for quadratic open
  {F}ermi systems},}\ }\href {\doibase 10.1088/1367-2630/10/4/043026}
  {\bibfield  {journal} {\bibinfo  {journal} {New J. Phys.}\ }\textbf {\bibinfo
  {volume} {10}},\ \bibinfo {pages} {043026} (\bibinfo {year}
  {2008})}\BibitemShut {NoStop}%
\bibitem [{\citenamefont {Prosen}\ and\ \citenamefont
  {{\v{Z}}nidari{\v{c}}}(2009)}]{Prosen2009}%
  \BibitemOpen
  \bibfield  {author} {\bibinfo {author} {\bibfnamefont {T.}~\bibnamefont
  {Prosen}}\ and\ \bibinfo {author} {\bibfnamefont {M.}~\bibnamefont
  {{\v{Z}}nidari{\v{c}}}},\ }\bibfield  {title} {\enquote {\bibinfo {title}
  {Matrix product simulations of non-equilibrium steady states of quantum spin
  chains},}\ }\href {\doibase 10.1088/1742-5468/2009/02/p02035} {\bibfield
  {journal} {\bibinfo  {journal} {J. Stat. Mech.}\ }\textbf {\bibinfo {volume}
  {2009}},\ \bibinfo {pages} {P02035} (\bibinfo {year} {2009})}\BibitemShut
  {NoStop}%
\bibitem [{\citenamefont
  {{\v{Z}}nidari{\v{c}}}(2010{\natexlab{a}})}]{Znidaric2010}%
  \BibitemOpen
  \bibfield  {author} {\bibinfo {author} {\bibfnamefont {M.}~\bibnamefont
  {{\v{Z}}nidari{\v{c}}}},\ }\bibfield  {title} {\enquote {\bibinfo {title}
  {Exact solution for a diffusive nonequilibrium steady state of an open
  quantum chain},}\ }\href {\doibase 10.1088/1742-5468/2010/05/l05002}
  {\bibfield  {journal} {\bibinfo  {journal} {J. Stat. Mech.}\ }\textbf
  {\bibinfo {volume} {2010}},\ \bibinfo {pages} {L05002} (\bibinfo {year}
  {2010}{\natexlab{a}})}\BibitemShut {NoStop}%
\bibitem [{\citenamefont
  {{\v{Z}}nidari{\v{c}}}(2011{\natexlab{a}})}]{Znidaric2011}%
  \BibitemOpen
  \bibfield  {author} {\bibinfo {author} {\bibfnamefont {M.}~\bibnamefont
  {{\v{Z}}nidari{\v{c}}}},\ }\bibfield  {title} {\enquote {\bibinfo {title}
  {Spin transport in a one-dimensional anisotropic {H}eisenberg model},}\
  }\href {\doibase 10.1103/physrevlett.106.220601} {\bibfield  {journal}
  {\bibinfo  {journal} {Phys. Rev. Lett.}\ }\textbf {\bibinfo {volume} {106}},\
  \bibinfo {pages} {220601} (\bibinfo {year} {2011}{\natexlab{a}})}\BibitemShut
  {NoStop}%
\bibitem [{\citenamefont {Prosen}(2011)}]{Prosen2011a}%
  \BibitemOpen
  \bibfield  {author} {\bibinfo {author} {\bibfnamefont {T.}~\bibnamefont
  {Prosen}},\ }\bibfield  {title} {\enquote {\bibinfo {title} {Open {{\it XXZ}}
  spin chain: Nonequilibrium steady state and a strict bound on ballistic
  transport},}\ }\href {\doibase 10.1103/physrevlett.106.217206} {\bibfield
  {journal} {\bibinfo  {journal} {Phys. Rev. Lett.}\ }\textbf {\bibinfo
  {volume} {106}},\ \bibinfo {pages} {217206} (\bibinfo {year}
  {2011})}\BibitemShut {NoStop}%
\bibitem [{\citenamefont {{\v{Z}}nidari{\v{c}}}\ \emph
  {et~al.}(2011)\citenamefont {{\v{Z}}nidari{\v{c}}}, \citenamefont
  {{\v{Z}}unkovi{\v{c}}},\ and\ \citenamefont {Prosen}}]{Znidaric2011a}%
  \BibitemOpen
  \bibfield  {author} {\bibinfo {author} {\bibfnamefont {M.}~\bibnamefont
  {{\v{Z}}nidari{\v{c}}}}, \bibinfo {author} {\bibfnamefont {B.}~\bibnamefont
  {{\v{Z}}unkovi{\v{c}}}}, \ and\ \bibinfo {author} {\bibfnamefont
  {T.}~\bibnamefont {Prosen}},\ }\bibfield  {title} {\enquote {\bibinfo {title}
  {Transport properties of a boundary-driven one-dimensional gas of spinless
  fermions},}\ }\href {\doibase 10.1103/physreve.84.051115} {\bibfield
  {journal} {\bibinfo  {journal} {Phys. Rev. E}\ }\textbf {\bibinfo {volume}
  {84}},\ \bibinfo {pages} {051115} (\bibinfo {year} {2011})}\BibitemShut
  {NoStop}%
\bibitem [{\citenamefont {Kos}\ and\ \citenamefont {Prosen}(2017)}]{Kos2017}%
  \BibitemOpen
  \bibfield  {author} {\bibinfo {author} {\bibfnamefont {P.}~\bibnamefont
  {Kos}}\ and\ \bibinfo {author} {\bibfnamefont {T.}~\bibnamefont {Prosen}},\
  }\bibfield  {title} {\enquote {\bibinfo {title} {Time-dependent correlation
  functions in open quadratic fermionic systems},}\ }\href {\doibase
  10.1088/1742-5468/aa9681} {\bibfield  {journal} {\bibinfo  {journal} {J.
  Stat. Mech.}\ }\textbf {\bibinfo {volume} {2017}},\ \bibinfo {pages} {123103}
  (\bibinfo {year} {2017})}\BibitemShut {NoStop}%
\bibitem [{\citenamefont {Prosen}\ and\ \citenamefont
  {Pi{\v{z}}orn}(2008)}]{Prosen2008quantum}%
  \BibitemOpen
  \bibfield  {author} {\bibinfo {author} {\bibfnamefont {T.}~\bibnamefont
  {Prosen}}\ and\ \bibinfo {author} {\bibfnamefont {I.}~\bibnamefont
  {Pi{\v{z}}orn}},\ }\bibfield  {title} {\enquote {\bibinfo {title} {Quantum
  phase transition in a far-from-equilibrium steady state of an {{\it XY}} spin
  chain},}\ }\href {\doibase 10.1103/physrevlett.101.105701} {\bibfield
  {journal} {\bibinfo  {journal} {Phys. Rev. Lett.}\ }\textbf {\bibinfo
  {volume} {101}},\ \bibinfo {pages} {105701} (\bibinfo {year}
  {2008})}\BibitemShut {NoStop}%
\bibitem [{\citenamefont
  {{\v{Z}}nidari{\v{c}}}(2011{\natexlab{b}})}]{Znidaric2011solvable}%
  \BibitemOpen
  \bibfield  {author} {\bibinfo {author} {\bibfnamefont {M.}~\bibnamefont
  {{\v{Z}}nidari{\v{c}}}},\ }\bibfield  {title} {\enquote {\bibinfo {title}
  {Solvable quantum nonequilibrium model exhibiting a phase transition and a
  matrix product representation},}\ }\href {\doibase
  10.1103/physreve.83.011108} {\bibfield  {journal} {\bibinfo  {journal} {Phys.
  Rev. E}\ }\textbf {\bibinfo {volume} {83}},\ \bibinfo {pages} {011108}
  (\bibinfo {year} {2011}{\natexlab{b}})}\BibitemShut {NoStop}%
\bibitem [{\citenamefont {Banchi}\ \emph {et~al.}(2014)\citenamefont {Banchi},
  \citenamefont {Giorda},\ and\ \citenamefont {Zanardi}}]{Banchi2014}%
  \BibitemOpen
  \bibfield  {author} {\bibinfo {author} {\bibfnamefont {L.}~\bibnamefont
  {Banchi}}, \bibinfo {author} {\bibfnamefont {P.}~\bibnamefont {Giorda}}, \
  and\ \bibinfo {author} {\bibfnamefont {P.}~\bibnamefont {Zanardi}},\
  }\bibfield  {title} {\enquote {\bibinfo {title} {Quantum information-geometry
  of dissipative quantum phase transitions},}\ }\href {\doibase
  10.1103/physreve.89.022102} {\bibfield  {journal} {\bibinfo  {journal} {Phys.
  Rev. E}\ }\textbf {\bibinfo {volume} {89}},\ \bibinfo {pages} {022102}
  (\bibinfo {year} {2014})}\BibitemShut {NoStop}%
\bibitem [{\citenamefont
  {{\v{Z}}nidari{\v{c}}}(2010{\natexlab{b}})}]{vznidarivc2010matrix}%
  \BibitemOpen
  \bibfield  {author} {\bibinfo {author} {\bibfnamefont {M.}~\bibnamefont
  {{\v{Z}}nidari{\v{c}}}},\ }\bibfield  {title} {\enquote {\bibinfo {title} {A
  matrix product solution for a nonequilibrium steady state of an {XX}
  chain},}\ }\href {\doibase 10.1088/1751-8113/43/41/415004} {\bibfield
  {journal} {\bibinfo  {journal} {J. Phys. A}\ }\textbf {\bibinfo {volume}
  {43}},\ \bibinfo {pages} {415004} (\bibinfo {year}
  {2010}{\natexlab{b}})}\BibitemShut {NoStop}%
\bibitem [{\citenamefont {Dutta}\ and\ \citenamefont {Cooper}()}]{Dutta2020}%
  \BibitemOpen
  \bibfield  {author} {\bibinfo {author} {\bibfnamefont {S.}~\bibnamefont
  {Dutta}}\ and\ \bibinfo {author} {\bibfnamefont {N.~R.}\ \bibnamefont
  {Cooper}},\ }\bibfield  {title} {\enquote {\bibinfo {title} {Long-range
  coherence and multiple steady states in a lossy qubit array},}\ }\href@noop
  {} {\ }\Eprint {http://arxiv.org/abs/arXiv:2004.07981} {arXiv:2004.07981}
  \BibitemShut {NoStop}%
\bibitem [{\citenamefont {Bu{\v{c}}a}\ and\ \citenamefont
  {Prosen}(2014)}]{Buca2014}%
  \BibitemOpen
  \bibfield  {author} {\bibinfo {author} {\bibfnamefont {B.}~\bibnamefont
  {Bu{\v{c}}a}}\ and\ \bibinfo {author} {\bibfnamefont {T.}~\bibnamefont
  {Prosen}},\ }\bibfield  {title} {\enquote {\bibinfo {title} {Exactly solvable
  counting statistics in open weakly coupled interacting spin systems},}\
  }\href {\doibase 10.1103/physrevlett.112.067201} {\bibfield  {journal}
  {\bibinfo  {journal} {Phys. Rev. Lett.}\ }\textbf {\bibinfo {volume} {112}},\
  \bibinfo {pages} {067201} (\bibinfo {year} {2014})}\BibitemShut {NoStop}%
\bibitem [{\citenamefont {Malo}\ \emph {et~al.}(2018)\citenamefont {Malo},
  \citenamefont {van Nieuwenburg}, \citenamefont {Fischer},\ and\ \citenamefont
  {Daley}}]{Malo2018}%
  \BibitemOpen
  \bibfield  {author} {\bibinfo {author} {\bibfnamefont {J.~Yago}\ \bibnamefont
  {Malo}}, \bibinfo {author} {\bibfnamefont {E.~P.~L.}\ \bibnamefont {van
  Nieuwenburg}}, \bibinfo {author} {\bibfnamefont {M.~H.}\ \bibnamefont
  {Fischer}}, \ and\ \bibinfo {author} {\bibfnamefont {A.~J.}\ \bibnamefont
  {Daley}},\ }\bibfield  {title} {\enquote {\bibinfo {title} {Particle
  statistics and lossy dynamics of ultracold atoms in optical lattices},}\
  }\href {\doibase 10.1103/physreva.97.053614} {\bibfield  {journal} {\bibinfo
  {journal} {Phys. Rev. A}\ }\textbf {\bibinfo {volume} {97}},\ \bibinfo
  {pages} {053614} (\bibinfo {year} {2018})}\BibitemShut {NoStop}%
\bibitem [{\citenamefont {Paredes}\ \emph {et~al.}(2004)\citenamefont
  {Paredes}, \citenamefont {Widera}, \citenamefont {Murg}, \citenamefont
  {Mandel}, \citenamefont {F{\"o}lling}, \citenamefont {Cirac}, \citenamefont
  {Shlyapnikov}, \citenamefont {H{\"a}nsch},\ and\ \citenamefont
  {Bloch}}]{Paredes2004}%
  \BibitemOpen
  \bibfield  {author} {\bibinfo {author} {\bibfnamefont {B.}~\bibnamefont
  {Paredes}}, \bibinfo {author} {\bibfnamefont {A.}~\bibnamefont {Widera}},
  \bibinfo {author} {\bibfnamefont {V.}~\bibnamefont {Murg}}, \bibinfo {author}
  {\bibfnamefont {O.}~\bibnamefont {Mandel}}, \bibinfo {author} {\bibfnamefont
  {S.}~\bibnamefont {F{\"o}lling}}, \bibinfo {author} {\bibfnamefont
  {I.}~\bibnamefont {Cirac}}, \bibinfo {author} {\bibfnamefont {G.~V.}\
  \bibnamefont {Shlyapnikov}}, \bibinfo {author} {\bibfnamefont {T.~W.}\
  \bibnamefont {H{\"a}nsch}}, \ and\ \bibinfo {author} {\bibfnamefont
  {I.}~\bibnamefont {Bloch}},\ }\bibfield  {title} {\enquote {\bibinfo {title}
  {Tonks{\textendash}{G}irardeau gas of ultracold atoms in an optical
  lattice},}\ }\href {\doibase 10.1038/nature02530} {\bibfield  {journal}
  {\bibinfo  {journal} {Nature (London)}\ }\textbf {\bibinfo {volume} {429}},\
  \bibinfo {pages} {277} (\bibinfo {year} {2004})}\BibitemShut {NoStop}%
\bibitem [{\citenamefont {St{\"o}ferle}\ \emph {et~al.}(2004)\citenamefont
  {St{\"o}ferle}, \citenamefont {Moritz}, \citenamefont {Schori}, \citenamefont
  {K{\"o}hl},\ and\ \citenamefont {Esslinger}}]{Stoeferle2004}%
  \BibitemOpen
  \bibfield  {author} {\bibinfo {author} {\bibfnamefont {T.}~\bibnamefont
  {St{\"o}ferle}}, \bibinfo {author} {\bibfnamefont {H.}~\bibnamefont
  {Moritz}}, \bibinfo {author} {\bibfnamefont {C.}~\bibnamefont {Schori}},
  \bibinfo {author} {\bibfnamefont {M.}~\bibnamefont {K{\"o}hl}}, \ and\
  \bibinfo {author} {\bibfnamefont {T.}~\bibnamefont {Esslinger}},\ }\bibfield
  {title} {\enquote {\bibinfo {title} {Transition from a strongly interacting
  {1D} superfluid to a {M}ott insulator},}\ }\href {\doibase
  10.1103/physrevlett.92.130403} {\bibfield  {journal} {\bibinfo  {journal}
  {Phys. Rev. Lett.}\ }\textbf {\bibinfo {volume} {92}},\ \bibinfo {pages}
  {130403} (\bibinfo {year} {2004})}\BibitemShut {NoStop}%
\bibitem [{\citenamefont {Preiss}\ \emph {et~al.}(2015)\citenamefont {Preiss},
  \citenamefont {Ma}, \citenamefont {Tai}, \citenamefont {Lukin}, \citenamefont
  {Rispoli}, \citenamefont {Zupancic}, \citenamefont {Lahini}, \citenamefont
  {Islam},\ and\ \citenamefont {Greiner}}]{Preiss2015}%
  \BibitemOpen
  \bibfield  {author} {\bibinfo {author} {\bibfnamefont {P.~M.}\ \bibnamefont
  {Preiss}}, \bibinfo {author} {\bibfnamefont {R.}~\bibnamefont {Ma}}, \bibinfo
  {author} {\bibfnamefont {M.~E.}\ \bibnamefont {Tai}}, \bibinfo {author}
  {\bibfnamefont {A.}~\bibnamefont {Lukin}}, \bibinfo {author} {\bibfnamefont
  {M.}~\bibnamefont {Rispoli}}, \bibinfo {author} {\bibfnamefont
  {P.}~\bibnamefont {Zupancic}}, \bibinfo {author} {\bibfnamefont
  {Y.}~\bibnamefont {Lahini}}, \bibinfo {author} {\bibfnamefont
  {R.}~\bibnamefont {Islam}}, \ and\ \bibinfo {author} {\bibfnamefont
  {M.}~\bibnamefont {Greiner}},\ }\bibfield  {title} {\enquote {\bibinfo
  {title} {Strongly correlated quantum walks in optical lattices},}\ }\href
  {\doibase 10.1126/science.1260364} {\bibfield  {journal} {\bibinfo  {journal}
  {Science}\ }\textbf {\bibinfo {volume} {347}},\ \bibinfo {pages} {1229}
  (\bibinfo {year} {2015})}\BibitemShut {NoStop}%
\bibitem [{\citenamefont {Barontini}\ \emph {et~al.}(2013)\citenamefont
  {Barontini}, \citenamefont {Labouvie}, \citenamefont {Stubenrauch},
  \citenamefont {Vogler}, \citenamefont {Guarrera},\ and\ \citenamefont
  {Ott}}]{Barontini2013}%
  \BibitemOpen
  \bibfield  {author} {\bibinfo {author} {\bibfnamefont {G.}~\bibnamefont
  {Barontini}}, \bibinfo {author} {\bibfnamefont {R.}~\bibnamefont {Labouvie}},
  \bibinfo {author} {\bibfnamefont {F.}~\bibnamefont {Stubenrauch}}, \bibinfo
  {author} {\bibfnamefont {A.}~\bibnamefont {Vogler}}, \bibinfo {author}
  {\bibfnamefont {V.}~\bibnamefont {Guarrera}}, \ and\ \bibinfo {author}
  {\bibfnamefont {H.}~\bibnamefont {Ott}},\ }\bibfield  {title} {\enquote
  {\bibinfo {title} {Controlling the dynamics of an open many-body quantum
  system with localized dissipation},}\ }\href {\doibase
  10.1103/physrevlett.110.035302} {\bibfield  {journal} {\bibinfo  {journal}
  {Phys. Rev. Lett.}\ }\textbf {\bibinfo {volume} {110}},\ \bibinfo {pages}
  {035302} (\bibinfo {year} {2013})}\BibitemShut {NoStop}%
\bibitem [{\citenamefont {Daley}(2014)}]{Daley2014}%
  \BibitemOpen
  \bibfield  {author} {\bibinfo {author} {\bibfnamefont {A.~J.}\ \bibnamefont
  {Daley}},\ }\bibfield  {title} {\enquote {\bibinfo {title} {Quantum
  trajectories and open many-body quantum systems},}\ }\href {\doibase
  10.1080/00018732.2014.933502} {\bibfield  {journal} {\bibinfo  {journal}
  {Adv. Phys.}\ }\textbf {\bibinfo {volume} {63}},\ \bibinfo {pages} {77}
  (\bibinfo {year} {2014})}\BibitemShut {NoStop}%
\bibitem [{\citenamefont {Lindblad}(1976)}]{Lindblad1976}%
  \BibitemOpen
  \bibfield  {author} {\bibinfo {author} {\bibfnamefont {G.}~\bibnamefont
  {Lindblad}},\ }\bibfield  {title} {\enquote {\bibinfo {title} {On the
  generators of quantum dynamical semigroups},}\ }\href {\doibase
  10.1007/bf01608499} {\bibfield  {journal} {\bibinfo  {journal} {Commun. Math.
  Phys.}\ }\textbf {\bibinfo {volume} {48}},\ \bibinfo {pages} {119} (\bibinfo
  {year} {1976})}\BibitemShut {NoStop}%
\bibitem [{\citenamefont {Gorini}\ \emph {et~al.}(1976)\citenamefont {Gorini},
  \citenamefont {Kossakowski},\ and\ \citenamefont {Sudarshan}}]{Gorini1976}%
  \BibitemOpen
  \bibfield  {author} {\bibinfo {author} {\bibfnamefont {V.}~\bibnamefont
  {Gorini}}, \bibinfo {author} {\bibfnamefont {A.}~\bibnamefont {Kossakowski}},
  \ and\ \bibinfo {author} {\bibfnamefont {E.~C.~G.}\ \bibnamefont
  {Sudarshan}},\ }\bibfield  {title} {\enquote {\bibinfo {title} {Completely
  positive dynamical semigroups of {$N$}-level systems},}\ }\href {\doibase
  10.1063/1.522979} {\bibfield  {journal} {\bibinfo  {journal} {J. Math.
  Phys.}\ }\textbf {\bibinfo {volume} {17}},\ \bibinfo {pages} {821} (\bibinfo
  {year} {1976})}\BibitemShut {NoStop}%
\bibitem [{\citenamefont {Breuer}\ and\ \citenamefont
  {Petruccione}(2002)}]{breuer2002theory}%
  \BibitemOpen
  \bibfield  {author} {\bibinfo {author} {\bibfnamefont {H.-P.}\ \bibnamefont
  {Breuer}}\ and\ \bibinfo {author} {\bibfnamefont {F.}~\bibnamefont
  {Petruccione}},\ }\href@noop {} {\emph {\bibinfo {title} {The Theory of Open
  Quantum Systems}}}\ (\bibinfo  {publisher} {Oxford University Press, Oxford,
  UK},\ \bibinfo {year} {2002})\BibitemShut {NoStop}%
\bibitem [{sup()}]{supplement}%
  \BibitemOpen
  \bibinfo {note} {See the Supplement, which includes Refs.~\cite{Prosen2012a,
  umezawa1995advanced, medvedyeva2016exact, karevski2013exact}, for analytic
  and perturbative solution for steady states, derivation of effective dynamics
  and rate equations at weak and strong dissipation, and examples of
  dissipation-induced long-range coherence in resonant ``dipole''
  geometries.}\BibitemShut {Stop}%
\bibitem [{\citenamefont {Pi{\v{z}}orn}(2013)}]{Pizorn2013}%
  \BibitemOpen
  \bibfield  {author} {\bibinfo {author} {\bibfnamefont {I.}~\bibnamefont
  {Pi{\v{z}}orn}},\ }\bibfield  {title} {\enquote {\bibinfo {title}
  {One-dimensional {B}ose-{H}ubbard model far from equilibrium},}\ }\href
  {\doibase 10.1103/physreva.88.043635} {\bibfield  {journal} {\bibinfo
  {journal} {Phys. Rev. A}\ }\textbf {\bibinfo {volume} {88}},\ \bibinfo
  {pages} {043635} (\bibinfo {year} {2013})}\BibitemShut {NoStop}%
\bibitem [{\citenamefont {Bu{\v{c}}a}\ and\ \citenamefont
  {Prosen}(2012)}]{Buca2012}%
  \BibitemOpen
  \bibfield  {author} {\bibinfo {author} {\bibfnamefont {B.}~\bibnamefont
  {Bu{\v{c}}a}}\ and\ \bibinfo {author} {\bibfnamefont {T.}~\bibnamefont
  {Prosen}},\ }\bibfield  {title} {\enquote {\bibinfo {title} {A note on
  symmetry reductions of the {L}indblad equation: transport in constrained open
  spin chains},}\ }\href {\doibase 10.1088/1367-2630/14/7/073007} {\bibfield
  {journal} {\bibinfo  {journal} {New J. Phys.}\ }\textbf {\bibinfo {volume}
  {14}},\ \bibinfo {pages} {073007} (\bibinfo {year} {2012})}\BibitemShut
  {NoStop}%
\bibitem [{\citenamefont {Popkov}\ \emph {et~al.}(2018)\citenamefont {Popkov},
  \citenamefont {Essink}, \citenamefont {Presilla},\ and\ \citenamefont
  {Sch{\"u}tz}}]{Popkov2018}%
  \BibitemOpen
  \bibfield  {author} {\bibinfo {author} {\bibfnamefont {V.}~\bibnamefont
  {Popkov}}, \bibinfo {author} {\bibfnamefont {S.}~\bibnamefont {Essink}},
  \bibinfo {author} {\bibfnamefont {C.}~\bibnamefont {Presilla}}, \ and\
  \bibinfo {author} {\bibfnamefont {G.}~\bibnamefont {Sch{\"u}tz}},\ }\bibfield
   {title} {\enquote {\bibinfo {title} {Effective quantum {Z}eno dynamics in
  dissipative quantum systems},}\ }\href {\doibase 10.1103/physreva.98.052110}
  {\bibfield  {journal} {\bibinfo  {journal} {Phys. Rev. A}\ }\textbf {\bibinfo
  {volume} {98}},\ \bibinfo {pages} {052110} (\bibinfo {year}
  {2018})}\BibitemShut {NoStop}%
\bibitem [{\citenamefont {Misra}\ and\ \citenamefont
  {Sudarshan}(1977)}]{misra1977zeno}%
  \BibitemOpen
  \bibfield  {author} {\bibinfo {author} {\bibfnamefont {B}~\bibnamefont
  {Misra}}\ and\ \bibinfo {author} {\bibfnamefont {E.~C.~G.}\ \bibnamefont
  {Sudarshan}},\ }\bibfield  {title} {\enquote {\bibinfo {title} {The {Z}eno's
  paradox in quantum theory},}\ }\href {\doibase 10.1063/1.523304} {\bibfield
  {journal} {\bibinfo  {journal} {J. Math. Phys.}\ }\textbf {\bibinfo {volume}
  {18}},\ \bibinfo {pages} {756} (\bibinfo {year} {1977})}\BibitemShut
  {NoStop}%
\bibitem [{\citenamefont {Carusotto}\ \emph {et~al.}(2020)\citenamefont
  {Carusotto}, \citenamefont {Houck}, \citenamefont {Koll{\'{a}}r},
  \citenamefont {Roushan}, \citenamefont {Schuster},\ and\ \citenamefont
  {Simon}}]{Carusotto2020}%
  \BibitemOpen
  \bibfield  {author} {\bibinfo {author} {\bibfnamefont {I.}~\bibnamefont
  {Carusotto}}, \bibinfo {author} {\bibfnamefont {A.~A.}\ \bibnamefont
  {Houck}}, \bibinfo {author} {\bibfnamefont {A.~J.}\ \bibnamefont
  {Koll{\'{a}}r}}, \bibinfo {author} {\bibfnamefont {P.}~\bibnamefont
  {Roushan}}, \bibinfo {author} {\bibfnamefont {D.~I.}\ \bibnamefont
  {Schuster}}, \ and\ \bibinfo {author} {\bibfnamefont {J.}~\bibnamefont
  {Simon}},\ }\bibfield  {title} {\enquote {\bibinfo {title} {Photonic
  materials in circuit quantum electrodynamics},}\ }\href {\doibase
  10.1038/s41567-020-0815-y} {\bibfield  {journal} {\bibinfo  {journal} {Nat.
  Phys.}\ }\textbf {\bibinfo {volume} {16}},\ \bibinfo {pages} {268} (\bibinfo
  {year} {2020})}\BibitemShut {NoStop}%
\bibitem [{\citenamefont {Popkov}\ \emph {et~al.}()\citenamefont {Popkov},
  \citenamefont {Essink}, \citenamefont {Kollath},\ and\ \citenamefont
  {Presilla}}]{Popkov2020}%
  \BibitemOpen
  \bibfield  {author} {\bibinfo {author} {\bibfnamefont {V.}~\bibnamefont
  {Popkov}}, \bibinfo {author} {\bibfnamefont {S.}~\bibnamefont {Essink}},
  \bibinfo {author} {\bibfnamefont {C.}~\bibnamefont {Kollath}}, \ and\
  \bibinfo {author} {\bibfnamefont {C.}~\bibnamefont {Presilla}},\ }\bibfield
  {title} {\enquote {\bibinfo {title} {Dissipative generation of pure steady
  states and a gambler ruin problem},}\ }\href@noop {} {\ }\Eprint
  {http://arxiv.org/abs/arXiv:2003.12149} {arXiv:2003.12149} \BibitemShut
  {NoStop}%
\bibitem [{\citenamefont {Prosen}\ and\ \citenamefont
  {{\v{Z}}nidari{\v{c}}}(2010)}]{Prosen2010}%
  \BibitemOpen
  \bibfield  {author} {\bibinfo {author} {\bibfnamefont {T.}~\bibnamefont
  {Prosen}}\ and\ \bibinfo {author} {\bibfnamefont {M.}~\bibnamefont
  {{\v{Z}}nidari{\v{c}}}},\ }\bibfield  {title} {\enquote {\bibinfo {title}
  {Long-range order in nonequilibrium interacting quantum spin chains},}\
  }\href {\doibase 10.1103/physrevlett.105.060603} {\bibfield  {journal}
  {\bibinfo  {journal} {Phys. Rev. Lett.}\ }\textbf {\bibinfo {volume} {105}},\
  \bibinfo {pages} {060603} (\bibinfo {year} {2010})}\BibitemShut {NoStop}%
\bibitem [{\citenamefont {Maghrebi}\ and\ \citenamefont
  {Gorshkov}(2016)}]{Maghrebi2016}%
  \BibitemOpen
  \bibfield  {author} {\bibinfo {author} {\bibfnamefont {M.~F.}\ \bibnamefont
  {Maghrebi}}\ and\ \bibinfo {author} {\bibfnamefont {A.~V.}\ \bibnamefont
  {Gorshkov}},\ }\bibfield  {title} {\enquote {\bibinfo {title} {Nonequilibrium
  many-body steady states via {K}eldysh formalism},}\ }\href {\doibase
  10.1103/physrevb.93.014307} {\bibfield  {journal} {\bibinfo  {journal} {Phys.
  Rev. B}\ }\textbf {\bibinfo {volume} {93}},\ \bibinfo {pages} {014307}
  (\bibinfo {year} {2016})}\BibitemShut {NoStop}%
\bibitem [{\citenamefont {Fradkin}(1989)}]{fradkin1989jordan}%
  \BibitemOpen
  \bibfield  {author} {\bibinfo {author} {\bibfnamefont {E.}~\bibnamefont
  {Fradkin}},\ }\bibfield  {title} {\enquote {\bibinfo {title}
  {{J}ordan-{W}igner transformation for quantum-spin systems in two dimensions
  and fractional statistics},}\ }\href {\doibase 10.1103/PhysRevLett.63.322}
  {\bibfield  {journal} {\bibinfo  {journal} {Phys. Rev. Lett.}\ }\textbf
  {\bibinfo {volume} {63}},\ \bibinfo {pages} {322} (\bibinfo {year}
  {1989})}\BibitemShut {NoStop}%
\bibitem [{\citenamefont {Umucal{\i}lar}\ and\ \citenamefont
  {Carusotto}(2017)}]{Umucalilar2017}%
  \BibitemOpen
  \bibfield  {author} {\bibinfo {author} {\bibfnamefont {R.~O.}\ \bibnamefont
  {Umucal{\i}lar}}\ and\ \bibinfo {author} {\bibfnamefont {I.}~\bibnamefont
  {Carusotto}},\ }\bibfield  {title} {\enquote {\bibinfo {title} {Generation
  and spectroscopic signatures of a fractional quantum {H}all liquid of photons
  in an incoherently pumped optical cavity},}\ }\href {\doibase
  10.1103/physreva.96.053808} {\bibfield  {journal} {\bibinfo  {journal} {Phys.
  Rev. A}\ }\textbf {\bibinfo {volume} {96}},\ \bibinfo {pages} {053808}
  (\bibinfo {year} {2017})}\BibitemShut {NoStop}%
\bibitem [{\citenamefont {Corman}\ \emph {et~al.}(2019)\citenamefont {Corman},
  \citenamefont {Fabritius}, \citenamefont {H{\"a}usler}, \citenamefont
  {Mohan}, \citenamefont {Dogra}, \citenamefont {Husmann}, \citenamefont
  {Lebrat},\ and\ \citenamefont {Esslinger}}]{Corman2019}%
  \BibitemOpen
  \bibfield  {author} {\bibinfo {author} {\bibfnamefont {L.}~\bibnamefont
  {Corman}}, \bibinfo {author} {\bibfnamefont {P.}~\bibnamefont {Fabritius}},
  \bibinfo {author} {\bibfnamefont {S.}~\bibnamefont {H{\"a}usler}}, \bibinfo
  {author} {\bibfnamefont {J.}~\bibnamefont {Mohan}}, \bibinfo {author}
  {\bibfnamefont {L.~H.}\ \bibnamefont {Dogra}}, \bibinfo {author}
  {\bibfnamefont {D.}~\bibnamefont {Husmann}}, \bibinfo {author} {\bibfnamefont
  {M.}~\bibnamefont {Lebrat}}, \ and\ \bibinfo {author} {\bibfnamefont
  {T.}~\bibnamefont {Esslinger}},\ }\bibfield  {title} {\enquote {\bibinfo
  {title} {Quantized conductance through a dissipative atomic point contact},}\
  }\href {\doibase 10.1103/physreva.100.053605} {\bibfield  {journal} {\bibinfo
   {journal} {Phys. Rev. A}\ }\textbf {\bibinfo {volume} {100}},\ \bibinfo
  {pages} {053605} (\bibinfo {year} {2019})}\BibitemShut {NoStop}%
\bibitem [{\citenamefont {Prosen}(2012)}]{Prosen2012a}%
  \BibitemOpen
  \bibfield  {author} {\bibinfo {author} {\bibfnamefont {T.}~\bibnamefont
  {Prosen}},\ }\bibfield  {title} {\enquote {\bibinfo {title} {Comments on a
  boundary-driven open {{\it XXZ}} chain: asymmetric driving and uniqueness of
  steady states},}\ }\href {\doibase 10.1088/0031-8949/86/05/058511} {\bibfield
   {journal} {\bibinfo  {journal} {Phys. Scr.}\ }\textbf {\bibinfo {volume}
  {86}},\ \bibinfo {pages} {058511} (\bibinfo {year} {2012})}\BibitemShut
  {NoStop}%
\bibitem [{\citenamefont {Umezawa}(1993)}]{umezawa1995advanced}%
  \BibitemOpen
  \bibfield  {author} {\bibinfo {author} {\bibfnamefont {H.}~\bibnamefont
  {Umezawa}},\ }\href@noop {} {\emph {\bibinfo {title} {Advanced Field Theory:
  Micro, Macro, and Thermal Physics}}}\ (\bibinfo  {publisher} {AIP Press, New
  York},\ \bibinfo {year} {1993})\BibitemShut {NoStop}%
\bibitem [{\citenamefont {Medvedyeva}\ \emph {et~al.}(2016)\citenamefont
  {Medvedyeva}, \citenamefont {Essler},\ and\ \citenamefont
  {Prosen}}]{medvedyeva2016exact}%
  \BibitemOpen
  \bibfield  {author} {\bibinfo {author} {\bibfnamefont {M.~V.}\ \bibnamefont
  {Medvedyeva}}, \bibinfo {author} {\bibfnamefont {F.~H.~L.}\ \bibnamefont
  {Essler}}, \ and\ \bibinfo {author} {\bibfnamefont {T.}~\bibnamefont
  {Prosen}},\ }\bibfield  {title} {\enquote {\bibinfo {title} {Exact {B}ethe
  ansatz spectrum of a tight-binding chain with dephasing noise},}\ }\href
  {\doibase 10.1103/PhysRevLett.117.137202} {\bibfield  {journal} {\bibinfo
  {journal} {Phys. Rev. Lett.}\ }\textbf {\bibinfo {volume} {117}},\ \bibinfo
  {pages} {137202} (\bibinfo {year} {2016})}\BibitemShut {NoStop}%
\bibitem [{\citenamefont {Karevski}\ \emph {et~al.}(2013)\citenamefont
  {Karevski}, \citenamefont {Popkov},\ and\ \citenamefont
  {Sch{\"u}tz}}]{karevski2013exact}%
  \BibitemOpen
  \bibfield  {author} {\bibinfo {author} {\bibfnamefont {D.}~\bibnamefont
  {Karevski}}, \bibinfo {author} {\bibfnamefont {V.}~\bibnamefont {Popkov}}, \
  and\ \bibinfo {author} {\bibfnamefont {G.~M.}\ \bibnamefont {Sch{\"u}tz}},\
  }\bibfield  {title} {\enquote {\bibinfo {title} {Exact matrix product
  solution for the boundary-driven {L}indblad {{\it XXZ}} chain},}\ }\href
  {\doibase 10.1103/PhysRevLett.110.047201} {\bibfield  {journal} {\bibinfo
  {journal} {Phys. Rev. Lett.}\ }\textbf {\bibinfo {volume} {110}},\ \bibinfo
  {pages} {047201} (\bibinfo {year} {2013})}\BibitemShut {NoStop}%
\end{thebibliography}

\begin{thebibliography}{8}%
\makeatletter
\providecommand \@ifxundefined [1]{%
 \@ifx{#1\undefined}
}%
\providecommand \@ifnum [1]{%
 \ifnum #1\expandafter \@firstoftwo
 \else \expandafter \@secondoftwo
 \fi
}%
\providecommand \@ifx [1]{%
 \ifx #1\expandafter \@firstoftwo
 \else \expandafter \@secondoftwo
 \fi
}%
\providecommand \natexlab [1]{#1}%
\providecommand \enquote  [1]{``#1''}%
\providecommand \bibnamefont  [1]{#1}%
\providecommand \bibfnamefont [1]{#1}%
\providecommand \citenamefont [1]{#1}%
\providecommand \href@noop [0]{\@secondoftwo}%
\providecommand \href [0]{\begingroup \@sanitize@url \@href}%
\providecommand \@href[1]{\@@startlink{#1}\@@href}%
\providecommand \@@href[1]{\endgroup#1\@@endlink}%
\providecommand \@sanitize@url [0]{\catcode `\\12\catcode `\$12\catcode
  `\&12\catcode `\#12\catcode `\^12\catcode `\_12\catcode `\%12\relax}%
\providecommand \@@startlink[1]{}%
\providecommand \@@endlink[0]{}%
\providecommand \url  [0]{\begingroup\@sanitize@url \@url }%
\providecommand \@url [1]{\endgroup\@href {#1}{\urlprefix }}%
\providecommand \urlprefix  [0]{URL }%
\providecommand \Eprint [0]{\href }%
\providecommand \doibase [0]{http://dx.doi.org/}%
\providecommand \selectlanguage [0]{\@gobble}%
\providecommand \bibinfo  [0]{\@secondoftwo}%
\providecommand \bibfield  [0]{\@secondoftwo}%
\providecommand \translation [1]{[#1]}%
\providecommand \BibitemOpen [0]{}%
\providecommand \bibitemStop [0]{}%
\providecommand \bibitemNoStop [0]{.\EOS\space}%
\providecommand \EOS [0]{\spacefactor3000\relax}%
\providecommand \BibitemShut  [1]{\csname bibitem#1\endcsname}%
\let\auto@bib@innerbib\@empty
%</preamble>
\bibitem [{\citenamefont {Prosen}(2008)}]{sProsen2008}%
  \BibitemOpen
  \bibfield  {author} {\bibinfo {author} {\bibfnamefont {T.}~\bibnamefont
  {Prosen}},\ }\bibfield  {title} {\enquote {\bibinfo {title} {Third
  quantization: a general method to solve master equations for quadratic open
  {F}ermi systems},}\ }\href {\doibase 10.1088/1367-2630/10/4/043026}
  {\bibfield  {journal} {\bibinfo  {journal} {New J. Phys.}\ }\textbf {\bibinfo
  {volume} {10}},\ \bibinfo {pages} {043026} (\bibinfo {year}
  {2008})}\BibitemShut {NoStop}%
\bibitem [{\citenamefont {Prosen}(2012)}]{sProsen2012a}%
  \BibitemOpen
  \bibfield  {author} {\bibinfo {author} {\bibfnamefont {T.}~\bibnamefont
  {Prosen}},\ }\bibfield  {title} {\enquote {\bibinfo {title} {Comments on a
  boundary-driven open {{\it XXZ}} chain: asymmetric driving and uniqueness of
  steady states},}\ }\href {\doibase 10.1088/0031-8949/86/05/058511} {\bibfield
   {journal} {\bibinfo  {journal} {Phys. Scr.}\ }\textbf {\bibinfo {volume}
  {86}},\ \bibinfo {pages} {058511} (\bibinfo {year} {2012})}\BibitemShut
  {NoStop}%
\bibitem [{\citenamefont {{\v{Z}}nidari{\v{c}}}(2010)}]{svznidarivc2010matrix}%
  \BibitemOpen
  \bibfield  {author} {\bibinfo {author} {\bibfnamefont {M.}~\bibnamefont
  {{\v{Z}}nidari{\v{c}}}},\ }\bibfield  {title} {\enquote {\bibinfo {title} {A
  matrix product solution for a nonequilibrium steady state of an {XX}
  chain},}\ }\href {\doibase 10.1088/1751-8113/43/41/415004} {\bibfield
  {journal} {\bibinfo  {journal} {J. Phys. A}\ }\textbf {\bibinfo {volume}
  {43}},\ \bibinfo {pages} {415004} (\bibinfo {year} {2010})}\BibitemShut
  {NoStop}%
\bibitem [{\citenamefont {Umezawa}(1993)}]{sumezawa1995advanced}%
  \BibitemOpen
  \bibfield  {author} {\bibinfo {author} {\bibfnamefont {H.}~\bibnamefont
  {Umezawa}},\ }\href@noop {} {\emph {\bibinfo {title} {Advanced Field Theory:
  Micro, Macro, and Thermal Physics}}}\ (\bibinfo  {publisher} {AIP Press, New
  York},\ \bibinfo {year} {1993})\BibitemShut {NoStop}%
\bibitem [{\citenamefont {Medvedyeva}\ \emph {et~al.}(2016)\citenamefont
  {Medvedyeva}, \citenamefont {Essler},\ and\ \citenamefont
  {Prosen}}]{smedvedyeva2016exact}%
  \BibitemOpen
  \bibfield  {author} {\bibinfo {author} {\bibfnamefont {M.~V.}\ \bibnamefont
  {Medvedyeva}}, \bibinfo {author} {\bibfnamefont {F.~H.~L.}\ \bibnamefont
  {Essler}}, \ and\ \bibinfo {author} {\bibfnamefont {T.}~\bibnamefont
  {Prosen}},\ }\bibfield  {title} {\enquote {\bibinfo {title} {Exact {B}ethe
  ansatz spectrum of a tight-binding chain with dephasing noise},}\ }\href
  {\doibase 10.1103/PhysRevLett.117.137202} {\bibfield  {journal} {\bibinfo
  {journal} {Phys. Rev. Lett.}\ }\textbf {\bibinfo {volume} {117}},\ \bibinfo
  {pages} {137202} (\bibinfo {year} {2016})}\BibitemShut {NoStop}%
\bibitem [{\citenamefont {Karevski}\ \emph {et~al.}(2013)\citenamefont
  {Karevski}, \citenamefont {Popkov},\ and\ \citenamefont
  {Sch{\"u}tz}}]{skarevski2013exact}%
  \BibitemOpen
  \bibfield  {author} {\bibinfo {author} {\bibfnamefont {D.}~\bibnamefont
  {Karevski}}, \bibinfo {author} {\bibfnamefont {V.}~\bibnamefont {Popkov}}, \
  and\ \bibinfo {author} {\bibfnamefont {G.~M.}\ \bibnamefont {Sch{\"u}tz}},\
  }\bibfield  {title} {\enquote {\bibinfo {title} {Exact matrix product
  solution for the boundary-driven {L}indblad {{\it XXZ}} chain},}\ }\href
  {\doibase 10.1103/PhysRevLett.110.047201} {\bibfield  {journal} {\bibinfo
  {journal} {Phys. Rev. Lett.}\ }\textbf {\bibinfo {volume} {110}},\ \bibinfo
  {pages} {047201} (\bibinfo {year} {2013})}\BibitemShut {NoStop}%
\bibitem [{\citenamefont {Daley}(2014)}]{sDaley2014}%
  \BibitemOpen
  \bibfield  {author} {\bibinfo {author} {\bibfnamefont {A.~J.}\ \bibnamefont
  {Daley}},\ }\bibfield  {title} {\enquote {\bibinfo {title} {Quantum
  trajectories and open many-body quantum systems},}\ }\href {\doibase
  10.1080/00018732.2014.933502} {\bibfield  {journal} {\bibinfo  {journal}
  {Adv. Phys.}\ }\textbf {\bibinfo {volume} {63}},\ \bibinfo {pages} {77}
  (\bibinfo {year} {2014})}\BibitemShut {NoStop}%
\bibitem [{\citenamefont {Popkov}\ \emph {et~al.}(2018)\citenamefont {Popkov},
  \citenamefont {Essink}, \citenamefont {Presilla},\ and\ \citenamefont
  {Sch{\"u}tz}}]{sPopkov2018}%
  \BibitemOpen
  \bibfield  {author} {\bibinfo {author} {\bibfnamefont {V.}~\bibnamefont
  {Popkov}}, \bibinfo {author} {\bibfnamefont {S.}~\bibnamefont {Essink}},
  \bibinfo {author} {\bibfnamefont {C.}~\bibnamefont {Presilla}}, \ and\
  \bibinfo {author} {\bibfnamefont {G.}~\bibnamefont {Sch{\"u}tz}},\ }\bibfield
   {title} {\enquote {\bibinfo {title} {Effective quantum {Z}eno dynamics in
  dissipative quantum systems},}\ }\href {\doibase 10.1103/physreva.98.052110}
  {\bibfield  {journal} {\bibinfo  {journal} {Phys. Rev. A}\ }\textbf {\bibinfo
  {volume} {98}},\ \bibinfo {pages} {052110} (\bibinfo {year}
  {2018})}\BibitemShut {NoStop}%
\end{thebibliography}


\end{document}